\documentclass[twocolumn,english]{revtex4-1}
\usepackage[T1]{fontenc}
\usepackage[latin9]{inputenc}
\usepackage{amsmath}
\usepackage{amssymb}
\usepackage{graphicx}
\usepackage{babel}

\usepackage{ulem}

\usepackage{color}

\begin{document}
\title{A Superradiant Maser with Nitrogen-Vacancy Center Spins}
\author{Qilong Wu}
\address{Henan Key Laboratory of Diamond Optoelectronic Materials and Devices, Key Laboratory of Material Physics Ministry of Education, School of Physics and Microelectronics, Zhengzhou University, Daxue Road 75, Zhengzhou 450052 China}
\author{Yuan Zhang}
\email{yzhuaudipc@dipc.org}
\address{Henan Key Laboratory of Diamond Optoelectronic Materials and Devices, Key Laboratory of Material Physics Ministry of Education, School of Physics and Microelectronics, Zhengzhou University, Daxue Road 75, Zhengzhou 450052 China}
\author{Xigui Yang}
\address{Henan Key Laboratory of Diamond Optoelectronic Materials and Devices, Key Laboratory of Material Physics Ministry of Education, School of Physics and Microelectronics, Zhengzhou University, Daxue Road 75, Zhengzhou 450052 China}

\author{Shi-Lei Su}
\address{School of Physics and Microelectronics, Zhengzhou University, Daxue Road 75, Zhengzhou 450052 China}

\author{Chongxin Shan}
\email{cxshan@zzu.edu.cn}
\address{Henan Key Laboratory of Diamond Optoelectronic Materials and Devices, Key Laboratory of Material Physics Ministry of Education, School of Physics and Microelectronics, Zhengzhou University, Daxue Road 75, Zhengzhou 450052 China}

\author{Klaus M{\o}lmer}
\email{moelmer@phys.au.dk}
\address{Center for Complex Quantum Systems, Department of Physics and Astronomy, Aarhus University, Ny Munkegade 120, DK-8000 Aarhus C, Denmark}

\begin{abstract}
Recent experiments have demonstrated Rabi-oscillations, superradiant pulses and stimulated emission from  negatively-charged nitrogen-vacancy ($\mathrm{NV}^{-}$) center spins in microwave resonators. These phenomena witness the kind of collective and strong coupling which has been prerequisite for observation of superradiant lasing in the optical frequency regime. In this article, we investigate the possibility to employ coherence, present in both the collective $\mathrm{NV}^{-}$ spin ensemble and the microwave field, to achieve a {\it superradiant maser}. Our calculations show that a superradiant maser with a linewidth below millihertz can be achieved with moderate kilohertz incoherent pumping of over $10^{14}$ spins kept at low temperature. We show that the superradiant masing prevails in the presence of inhomogeneous broadening, and we present numerical and analytical studies of the dependence of the phenomenon on the various physical parameters.
\end{abstract}

\maketitle
\section{Introduction}

\begin{figure}[htbp]
\centering \includegraphics[scale=0.4]{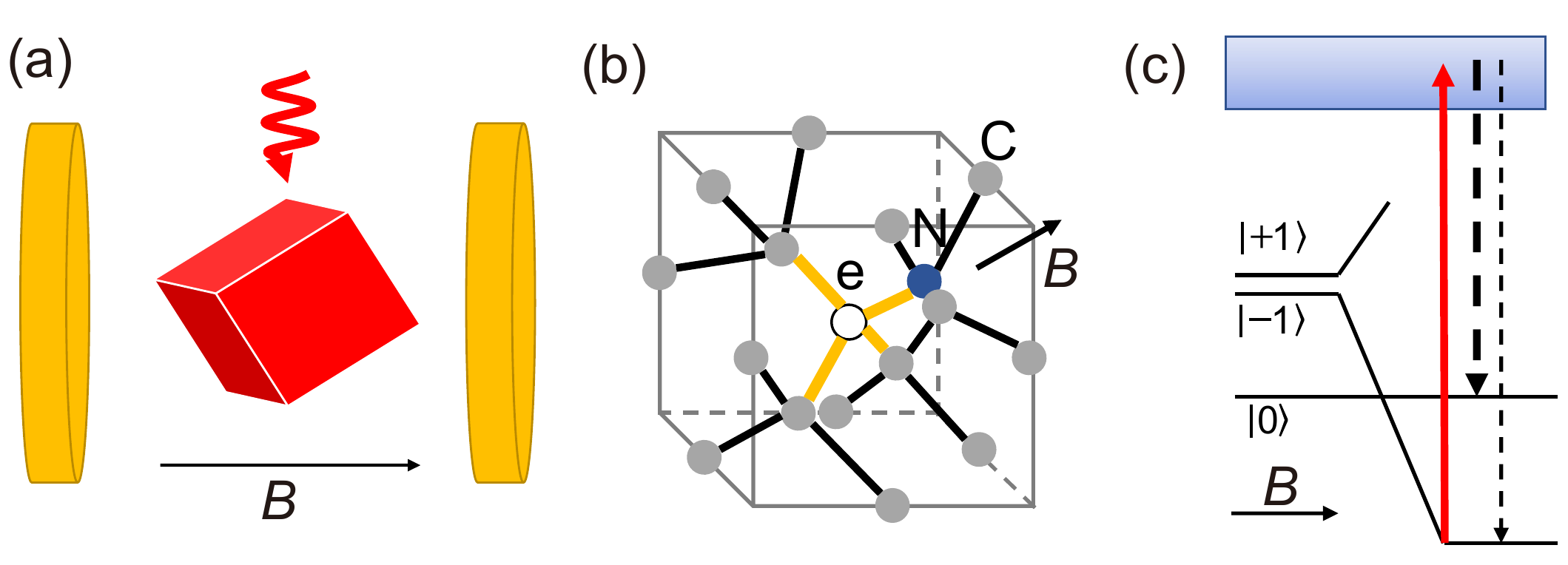}
\caption{\label{fig:systemlevel} Schematic of a superradiant maser with $\mathrm{NV}^{-}$ center spins. Panel (a) depicts a diamond  illuminated by laser light and subjected to a static magnetic field $B$ inside a microwave resonator. Panel (b) shows a negatively charged $\mathrm{NV}^{-}$ center formed by a nitrogen atom (blue sphere) and a carbon vacancy (empty sphere) in the lattice of carbon atoms (gray spheres). Panel (c) shows the simplified energy level structure of the $\mathrm{NV}^{-}$ center with electron-spin levels $\left|0\right\rangle$ and  $\left|\pm1\right\rangle $, split by the  magnetic field, and electronically excited states accessible by optical excitation (solid red arrow) from $\left|-1\right\rangle $, followed by radiative and non-radiative decay back to the ground states (dashed arrows). The latter processes favor  decay into the state $\left| 0\right\rangle $ and lead to population inversion between this state and the state $\left|-1\right\rangle $.}
\end{figure}

Room-temperature maser operation \cite{JDBreeze} was recently demonstrated with negatively-charged nitrogen-vacancy ($\mathrm{NV}^{-}$) center spins in diamond coupled to a microwave resonator, see the schematics of Fig. \ref{fig:systemlevel} (a). This observation was possible because the diamond host offers mechanical and thermal stability \cite{LJin,JDBreeze} so that the $\mathrm{NV}^{-}$ spins with long coherence time can be pumped through optical means. At cryogenic temperatures, the $\mathrm{NV}^{-}$ spin-resonator systems have also been used to demonstrate microwave superradiant pulses \cite{AAngerer}, coherent Rabi-oscillations \cite{SPutz}, and Rabi-splitting of microwave transmission spectra  \cite{RAmsuss,YKubo,AAngerer1},
enabled by the collective coupling between the  spins and the microwave resonator mode. 

Theoretical \cite{DMeiser,DMeiser1} and experiment studies \cite{MANorciaSciAdv,JGBohnet} have shown that steady-state superradiance can be achieved in the optical frequency domain when incoherently pumped atoms couple collectively but weakly with an optical cavity. This system benefits from the atomic coherence, and can be used  to achieve coherent light source with ultra-narrow linewidth, set by the Purcell enhanced atomic decay rate $\Gamma_c$ \cite{DMeiser,DMeiser1}. Furthermore, in the collective strong coupling regime, superradiant lasing can occur, which benefits from  both the photonic and atomic coherence, and has a linewidth even below $\Gamma_c$ \cite{DATieri,KDebnath,MANorcia}. 

The similarity of the physical mechanisms in the microwave and optical domain suggest that a superradiant maser may be achieved with the $\mathrm{NV}^{-}$ spin-resonator systems, which are currently subject to intense experimental studies. In this article we investigate the conditions needed to observe this phenomenon and we make quantitative predictions for its performance. Our studies show that a superradiant maser with linewidth down to millihertz can be achieved if a sufficiently large $\mathrm{NV}^{-}$ spin ensemble is pumped at a rate of kilohertz. This result applies even when the spin ensemble has a megahertz wide inhomogeneous broadening.

The article is organized as follows. In Sec. \ref{sec:sytem}, we introduce the $\mathrm{NV}^{-}$ spin-resonator system, the spin level scheme and the pumping mechanism. In Sec. \ref{sec:meq}, we present the quantum master equation for the system dynamics, solutions based on a second-order mean-field approach, and a method to compute the steady-state spectrum. In Sec. \ref{sec:res}, we apply our numerical methods and analytical approximations to analyze the dependence of the steady-states of the spin ensemble, by means of Dicke states, the intra-cavity photon number and the spectrum linewidth on the  number of $\mathrm{NV}^{-}$ spins, the pumping rate, the temperature and the spin-resonator frequency detuning. In the last section, we summarize our main conclusions. 

\section{The $\mathrm{NV}^{-}$ spin-resonator system  \label{sec:sytem}}

We consider the system sketched in Fig.  \ref{fig:systemlevel} (a), where a diamond inside a microwave resonator is subject to a static magnetic field and incoherent optical pumping. The schematics of the resonator may represent, e.g., a cylindrical copper cavity \cite{JDBreeze}, a 3D lumped element resonator \cite{AAngerer}, or a superconducting co-planar wave-guide resonator \cite{SPutz}. The geometric structure of the negatively-charged $\mathrm{NV}^{-}$ centers is shown in Fig. \ref{fig:systemlevel} (b). By aligning properly the magnetic field with one of four possible quantization axes (along the direction between the vacancy and the adjacent nitrogen atom), it will introduce large level-shifts to the corresponding spin eigenstates. The energy levels of a single $\mathrm{NV}^{-}$ center can then be simplified as shown in  Fig. \ref{fig:systemlevel} (c).

The spin state  $\left|0\right\rangle$ is not affected by the magnetic field, while the spin states $\left|+1\right\rangle$ and $\left|-1\right\rangle$ are shifted upwards and downwards, respectively.  By optically exciting the $\mathrm{NV}^{-}$ centers to the electronic excited state (red arrow) and utilizing the spin-sensitive, radiative and non-radiative decay mechanism from this state to the electronic ground state (dashed arrows), we can effectively pump the $\mathrm{NV}^{-}$ spins from the $\left|-1\right\rangle$ state to $\left|0\right\rangle$ state and create a population inversion between these spin states, which couples collectively with the microwave resonator field to form the superradiant maser. In addition, due to the nuclear spin of the nitrogen atom, and its hyperfine interaction with the electron spin, the energy levels become more involved and more transitions between the electron-nuclear spin levels become possible. However, for large magnetic field, only one of the transitions is resonant with the resonator mode. In this case, the states $\left|-1\right\rangle$ and $\left|0\right\rangle$ can be interpreted as the states of this resonant transition. For a more precise account of the $\mathrm{NV}^{-}$ spin levels, see the Appendix A.

\section{Master equation and mean-field solutions \label{sec:meq}}

To simplify the notation, we describe the two most-relevant spin states $\left|0\right\rangle$ and $\left|-1\right\rangle$ as the up and down components of a pseudo 1/2-spin, and we model the pumping by optical excitation and optical decay by an effective transfer rate from the state $\left|-1\right\rangle$ to $\left|0\right\rangle$. While this treatment of the pumping is often adopted in laser theories \cite{DMeiser,KDebnath}, it can be readily extended to include intermediate states \cite{YZhang1,YZhang2}. Accounting for the dissipation of the microwave resonator and the $\mathrm{NV}^{-}$ spins, we obtain the following master equation for the reduced density operator $\hat{\rho}$  of the system: 
\begin{align}
 & \frac{\partial}{\partial t}\hat{\rho}=-\frac{i}{\hbar}\left[\hat{H}_{s}+\hat{H}_{c}+\hat{H}_{s-c},\hat{\rho}\right]\nonumber \\
 & -\kappa_{c}\left\{ \left(n_{c}^{th}+1\right)\mathcal{D}\left[\hat{a}\right]\hat{\rho}+n_{c}^{th}\mathcal{D}\left[\hat{a}^{\dagger}\right]\hat{\rho}\right\} \nonumber \\
 & -\sum_{k}\gamma_{k}\left\{ \left(n_{k}^{th}+1\right)\mathcal{D}\left[\hat{\sigma}_{k}^{-}\right]\hat{\rho}+n_{k}^{th}\mathcal{D}\left[\hat{\sigma}_{k}^{\dagger}\right]\hat{\rho}\right\} \nonumber \\
 & -\sum_{k}\Bigl\{\eta_{k}\mathcal{D}\left[\hat{\sigma}_{k}^{\dagger}\right]\hat{\rho}+\frac{1}{2}\chi_{k}\mathcal{D}\left[\hat{\sigma}_{k}^{z}\right]\hat{\rho}\Bigr\}.\label{eq:master-equation}
\end{align}
Here, the Hamiltonian $\hat{H}_{s}=\hbar\sum_{k=1}^{N}\left(\omega_{k}/2\right)\hat{\sigma}_{k}^{z}$
includes the individual frequencies $\omega_{k}$
and Pauli operators $\hat{\sigma}_{k}^{z}$ for all the $N$ pseudo spins. The Hamiltonian $\hat{H}_{c}=\hbar\omega_{c}\hat{a}^{\dagger}\hat{a}$ specifies the frequency $\omega_{c}$,
the creation $\hat{a}^{\dagger}$ and annihilation operator $\hat{a}$ of
photons in the microwave resonator. The spin-resonator mode interaction $\hat{H}_{s-c}=\hbar\sum_{k}g_{k}\left(\hat{\sigma}_{k}^{\dagger}\hat{a}+\hat{a}^{\dagger}\hat{\sigma}_{k}^{-}\right)$ is given by the coupling strengths $g_{k}$, the raising $\hat{\sigma}_{k}^{\dagger}$
and lowering $\hat{\sigma}_{k}^{-}$ ladder operators for the pseudo spins.
The Lindblad relaxation terms in the second line of Eq. (\ref{eq:master-equation})
describe the exchange of microwave photons with rate $\kappa_c$ between the resonator and the surrounding thermal heat bath with the temperature $T$ and mean thermal excitation number $n_{c}^{th}=\left[e^{\hbar\omega_{c}/k_{B}T}-1\right]^{-1}$ ($k_{B}$ is the Boltzmann constant). The Lindblad terms in the third line of Eq. (\ref{eq:master-equation}) describe the similar thermal processes for the $\mathrm{NV}^{-}$ spins with the rate $\gamma_{k}$ and the mean thermal occupation number  $n_{k}^{th}=\left[e^{\hbar\omega_{k}/k_{B}T}-1\right]^{-1}$. The Lindblad
terms in the last line of Eq. (\ref{eq:master-equation}) describe
incoherent pumping with rates $\eta_{k}$, and dephasing with rates $\chi_{k}$ of the pseudo spins. In all the relaxation terms, the superoperator, for any operator $\hat{o}$, is defined as $\mathcal{D}\left[\hat{o}\right]\hat{\rho}=\frac{1}{2}\left\{ \hat{o}^{\dagger}\hat{o},\hat{\rho}\right\} -\hat{o}\hat{\rho} \hat{o}^{\dagger}$. 


The more than a trillion $\mathrm{NV}^{-}$ center spins encountered in the experiments \cite{JDBreeze,AAngerer,SPutz,RAmsuss,YKubo,AAngerer1}, preclude solution of the master equation Eq. \eqref{eq:master-equation} by standard density matrix techniques. Thus, in this article, we adopt a second-order mean-field approach \cite{KDebnath,DPlankensteiner} (equivalent to the cluster expansion methods \cite{HAMLeymann}), to solve the master equation. This method captures quantum correlations (see below), which are essential for the steady-state superradiance \cite{DMeiser,DMeiser1} and the superradiant lasing phenomena \cite{KDebnath}.  We use the master equation Eq. \eqref{eq:master-equation} to derive the equation $\partial_{t}\left\langle \hat{o} \right\rangle =\mathrm{tr}\left\{ \hat{o} \partial_{t}\hat{\rho}\right\} $
for the expectation value $\left\langle \hat{o} \right\rangle =\mathrm{tr}\left\{ \hat{o} \hat{\rho}\right\} $
of any observable $\hat{o}$. By setting $\hat{o}=\hat{a}^\dagger \hat{a}$, we, e.g., obtain the equation
for the intra-resonator photon number: 
\begin{align}
 & \frac{\partial}{\partial t}\left\langle \hat{a}^{\dagger}\hat{a}\right\rangle =-\kappa_{c}\left\langle \hat{a}^{\dagger}\hat{a}\right\rangle +\kappa_{c}n_{c}^{th}\nonumber \\
 & +i\sum_{k}g_{k}\left(\left\langle \hat{\sigma}_{k}^{\dagger}\hat{a}\right\rangle -\left\langle \hat{a}^{\dagger}\hat{\sigma}_{k}^{-}\right\rangle \right). \label{eq:apa}
\end{align}
The mean photon number is coupled to the spin-photon correlations $\left\langle \hat{\sigma}_{k}^{\dagger}\hat{a}\right\rangle$ (and their conjugation
$\left\langle \hat{a}^{\dagger}\hat{\sigma}_{k}^{-}\right\rangle $),
which, in turn, follow 
\begin{align}
 & \frac{\partial}{\partial t}\left\langle \hat{\sigma}_{k}^{\dagger}\hat{a}\right\rangle =i\left(\tilde{\omega}_{k}^{*}-\tilde{\omega}_{c}\right)\left\langle \hat{\sigma}_{k}^{\dagger}\hat{a}\right\rangle -ig_{k}\left\langle \hat{a}^{\dagger}\hat{a}\hat{\sigma}_{k}^{z}\right\rangle \nonumber \\
 & -i\frac{g_{k}}{2}\left(\left\langle \hat{\sigma}_{k}^{z}\right\rangle +1\right)-i\sum_{k'\ne k}g_{k'}\left\langle \hat{\sigma}_{k}^{\dagger}\hat{\sigma}_{k'}^{-}\right\rangle . \label{eq:spa}
\end{align}
Here, we have introduced the complex frequencies $\tilde{\omega}_{c}=\omega_{c}-i\kappa_{c}/2$
and $\tilde{\omega}_{k}=\omega_{k}-i\lambda_{k}^{s}$ with the total pseudo-spin
dephasing rate,  $\lambda_{k}^{s}=\frac{1}{2}\left[\gamma_{k}\left(2n_{k}^{th}+1\right)+\eta_{k}\right]+\chi_{k}$.
We note that the above equation includes a  third-order term $\left\langle \hat{a}^{\dagger}\hat{a}\hat{\sigma}_{k}^{z}\right\rangle $
which will evolve in a manner that depends on even higher-order terms. To truncate this hierarchy of equations, we assume the vanishing third-order cumulants \cite{DMeiser,DPlankensteiner,ARoth} and thus approximate third-order quantities, e.g. $\left\langle \hat{a}^{\dagger}\hat{a}\hat{\sigma}_{k}^{z}\right\rangle $, by the low-order expressions, e.g. $\left\langle \hat{a}^{\dagger}\right\rangle \left\langle \hat{a}\hat{\sigma}_{k}^{z}\right\rangle +\left\langle \hat{a}\right\rangle \left\langle \hat{a}^{\dagger}\hat{\sigma}_{k}^{z}\right\rangle +\left\langle \hat{\sigma}_{k}^{z}\right\rangle \left\langle \hat{a}^{\dagger}\hat{a}\right\rangle -2\left\langle \hat{a}^{\dagger}\right\rangle \left\langle \hat{a} \right\rangle \left\langle \hat{\sigma}_{k}^{z}\right\rangle $.
For the current system, the resonator field amplitude
$\left\langle \hat{a}\right\rangle $ and its conjugation $\left\langle  \hat{a}^{\dagger} \right\rangle $ are always zero and thus we have $\left\langle \hat{a}^{\dagger}\hat{a}\hat{\sigma}_{k}^{z}\right\rangle \approx\left\langle \hat{a}^{\dagger}\hat{a}\right\rangle \left\langle \hat{\sigma}_{k}^{z}\right\rangle $.
The equations for other terms, like the population difference $\left\langle \hat{\sigma}_{k}^{z}\right\rangle $ and the spin-spin correlations $\left\langle \hat{\sigma}_{k}^{\dagger}\hat{\sigma}_{k'}^{-}\right\rangle $, are detailed in the Appendix B.


To compute the spectrum of emission from the resonator, we mimic its experimental measurement
by coupling the field to a filter resonator \cite{KDebnath,YZhang-1}.
To this end, we complement the master equation (\ref{eq:master-equation})
with the additional terms $\frac{\partial}{\partial t}\hat{\rho}\propto-\frac{i}{\hbar}\left[\hat{H}_{f}+\hat{H}_{f-c},\hat{\rho}\right]-\kappa_{f}\mathcal{D}\left[\hat{b}\right]\hat{\rho} $. Here, the Hamiltonian $\hat{H}_{f}=\hbar\omega_{f}\hat{b}^{\dagger}\hat{b}$ describes the
filter resonator with the microwave frequency $\omega_{f}$, the creation
$\hat{b}^{\dagger}$ and annihilation operator $\hat{b}$, the Hamiltonian $\hat{H}_{f-c}=\hbar G\left(\hat{a}^{\dagger}\hat{b}+\hat{b}^{\dagger}\hat{a}\right)$
specifies the filter resonator-main resonator coupling with the strength
$G$, and the Lindblad term describes photon loss within the filter
resonator with a rate $\kappa_{f}$. Here, we consider a filter resonator at low temperature to properly define the emission spectrum of the system. Using again the second-order mean-field approach, we obtain the equation for the mean number of photons in the filter resonator 
\begin{align}
 & \frac{\partial}{\partial t}\left\langle \hat{b}^{\dagger}\hat{b}\right\rangle =iG\left(\left\langle \hat{a}^{\dagger}\hat{b}\right\rangle -\left\langle \hat{b}^{\dagger}\hat{a}\right\rangle \right)-\kappa_{f}\left\langle \hat{b}^{\dagger}\hat{b}\right\rangle,\label{eq:bpb}
\end{align}
which depends on the photon-photon correlation between the main resonator
and the filter resonator $\left\langle \hat{a}^{\dagger}\hat{b}\right\rangle $ (and $\left\langle \hat{b}^{\dagger}\hat{a}\right\rangle $). The equation for
this correlation and other terms like the spin-filter resonator correlation $\left\langle \hat{\sigma}_{k}^{\dagger}\hat{b}\right\rangle $ are given in the Appendix B.

\section{\label{sec:res}  Numerical results}

In this section we present numerical solutions of the coupled second order mean field equations and analytical approximations that characterize their behavior. While the method is general, we consider physical parameters compatible with the experiments described in Ref. \cite{JDBreeze}. The microwave mode of the resonator has a frequency of $\omega_{c}=2\pi\times9.22$ GHz, a cavity loss rate \textbf{ $\kappa_{c}=1.9$ }MHz. The spin ensemble contains about $N=4\times10^{13}$ $\mathrm{NV}^{-}$ centers, and has an inhomogeneously broadened distribution of transition frequencies $\omega_{k}\approx\omega_{c}$ of $4$ MHz. The individual spins couple with the resonator with the strength $g_{k}=0.7$ Hz, and their relaxation rate $\gamma_{k}\approx 0.157$ Hz can be extracted from the typical temperature-dependent spin-relaxation rate $(2n_k^{th}+1)\gamma_k\approx0.21$ kHz at room temperature $T=293$ K \cite{JDBreeze} (with the thermal excitation of spins $n_k^{th}\approx662$). Note that the longitudinal  spin-relaxation rate might vary largely across samples due to different mechanisms \cite{AJarmola,TAstner}.
As a result, the Purcell rate of spin relaxation via the resonator mode can be estimated as $\Gamma_c = 4g_{k}^2/\kappa_c \approx 1.03\times 10^{-6}$ Hz. Since the collective coupling,  $\sqrt{N}g_{k}=4.42$ MHz, is stronger than the loss rates, $\kappa_{c},\gamma_k,\chi_k$, the system allows exploration of the superradiant maser regime. For reference, the key parameters for other experimental setups \cite{AAngerer,SPutz,RAmsuss,YKubo,AAngerer1} are summarized in the Appendix C. 

In Subsec. \ref{sec:InhomBrodening}, we model the inhomogeneous distribution by assigning all spins to 50 different frequency classes, and assuming that the spin mean values (correlations) are identical for spins belonging to the same frequency class (pair of frequency classes) \citep{KDebnath1,ABychek}. The numerical calculations with this model will demonstrate the emission of radiation with a linewidth narrower than both the resonator linewidth and the inhomogeneous frequency profile of the spin ensemble, and it serves as reference for subsequent analyses in Subsec. \ref{sec:HomBrodening}, where we shall model the system inhomogeneity by an even simpler ansatz where the spins are identical, but subject to homogeneous broadening by a dephasing rate of 4 MHz. This simpler model permits exhaustive numerical studies of the dependence on different parameters, analytical approximations and explanations of the results.  

\begin{figure}[htbp]
\centering \includegraphics[scale=0.3]{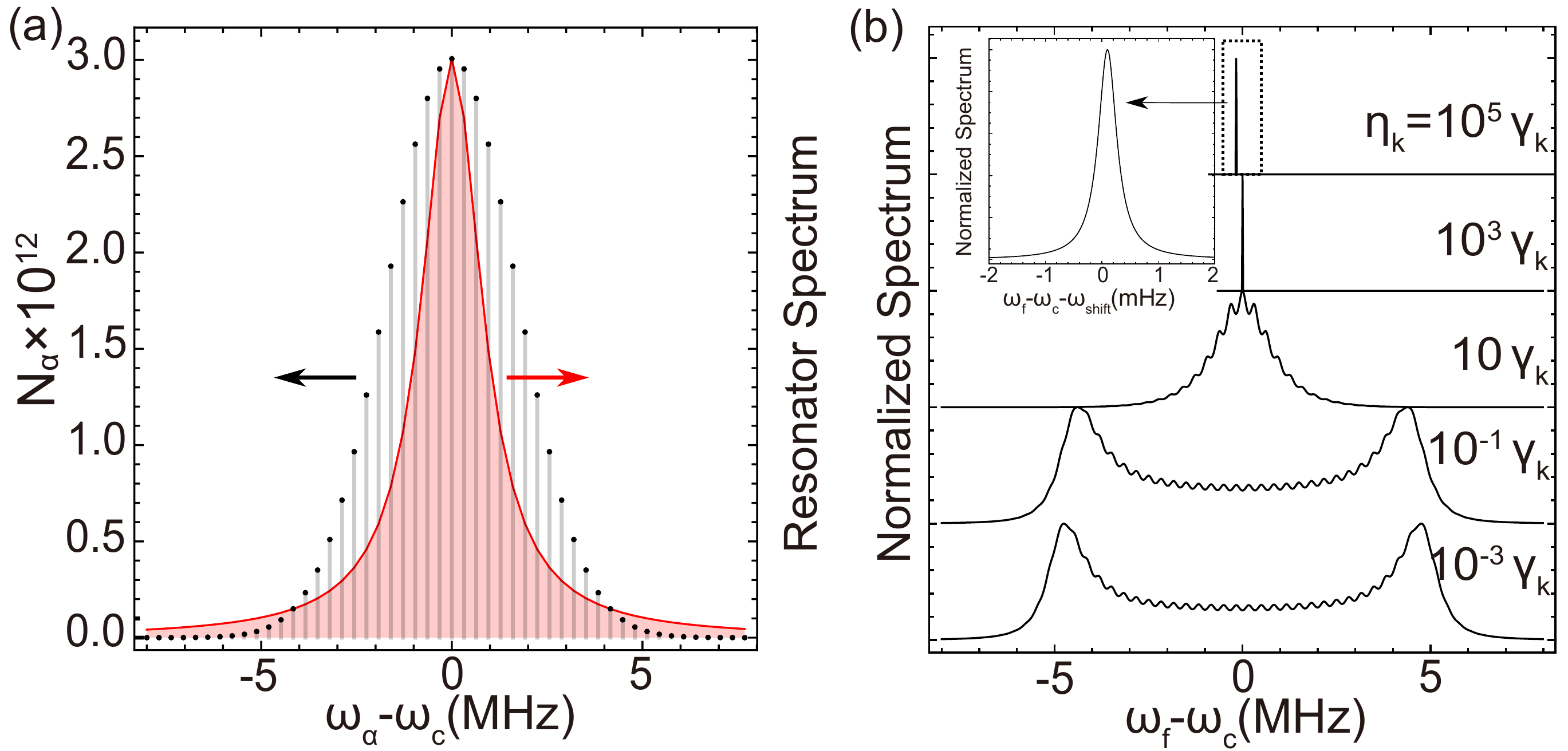} 
\caption{Superradiant masing of spin sub-ensembles. Panel (a) shows a $\chi_{inh}=$ 4 MHz wide Gaussian distribution of spin transition frequencies, modelled by assigning $N_\alpha$ spins to $N_{se}=50$ discrete  transition frequencies (thin grey vertical lines, left axis), where the total number of spins is $\sum_\alpha N_{\alpha} = 4\times 10^{13}$. The Lorentzian spectrum of the empty resonator is shown by the red curve and shaded area (arbitrary units, right axis). To represent the widths of the frequency intervals, we assume a dephasing rate of all individual spins of $\chi_k =2\chi_{inh}/N_{se}$. Panel (b) shows the calculated spectrum for different values of the pumping rate  $\eta_k$ between $10^{-3}\gamma_k$ and $10^5 \gamma_k$ (from bottom to upper curve), and the zoom-in of the sharp peak for $10^5 \gamma_k$ ($\omega_{shift}=211603.159$ Hz). The temperature is $25$ mK, and other parameters are specified in the main text.} 
\label{fig:gaussian} 
\end{figure}

\subsection{Modelling of the Dynamics and Emission Spectra of Spin Sub-ensembles \label{sec:InhomBrodening}}

$\mathrm{NV}^{-}$ spins in diamond suffer from  inhomogeneous broadening of a few MHz. For many purposes, this broadening may be treated with a single homogeneous dephasing rate $\chi_{k}$ \cite{LJin,AAngerer}, but for our calculation of an ultranarrow emission we want to make sure that our results are indeed reproducible with a model accounting for the actual variation of transition frequencies in the system. To this end, we will treat the total spin ensemble as collection of sub-ensembles labeled by $\alpha$, and assume that the number of spins $N_{\alpha}$ in each sub-ensemble represents a coarse grained approximation to the actual distribution \citep{KDebnath1,ABychek}. In our numerical calculations we assume a Gaussian frequency distribution, see Fig. \ref{fig:gaussian}(a). 

Furthermore, we assume that the spins in each sub-ensemble are identical, which allows us to reduce the mean-field equations in the Sec. \ref{sec:meq} to the much fewer equations for quantities related to individual sub-ensembles or between sub-ensembles, see the Appendix D. By solving these equations, we have computed the steady-state spectra for the system with $\sum_{\alpha} N_{\alpha} = 4\times 10^{13}$ spins distributed into $50$ sub-ensembles. Fig. \ref{fig:gaussian}(b) shows results for different values of the pumping rate $\eta_k$ (in units of the spin relaxation rate $\gamma_k$). For $\eta_k$ smaller than $\gamma_k$, we observe two broad peaks, while for moderate $\eta_k$ we obtain a single sharp peak. For the largest pumping rate studied, the linewidth of the sharp emission peak is around millihertz (see the inset). This sharp emission persists even if we split the center sub-ensemble into many subsubensembles about mHz wide spread. We attribute this robustness to the synchronization effect of spin sub-ensembles \cite{MXu,AShankar,KDebnath1}, and justify it by showing that the narrow spectrum is contributed mostly by the spin subensembles nearly degenerate with the resonator mode. For more details, see the Appendix D.

\begin{figure*}[htbp]
\centering \includegraphics[scale=0.3]{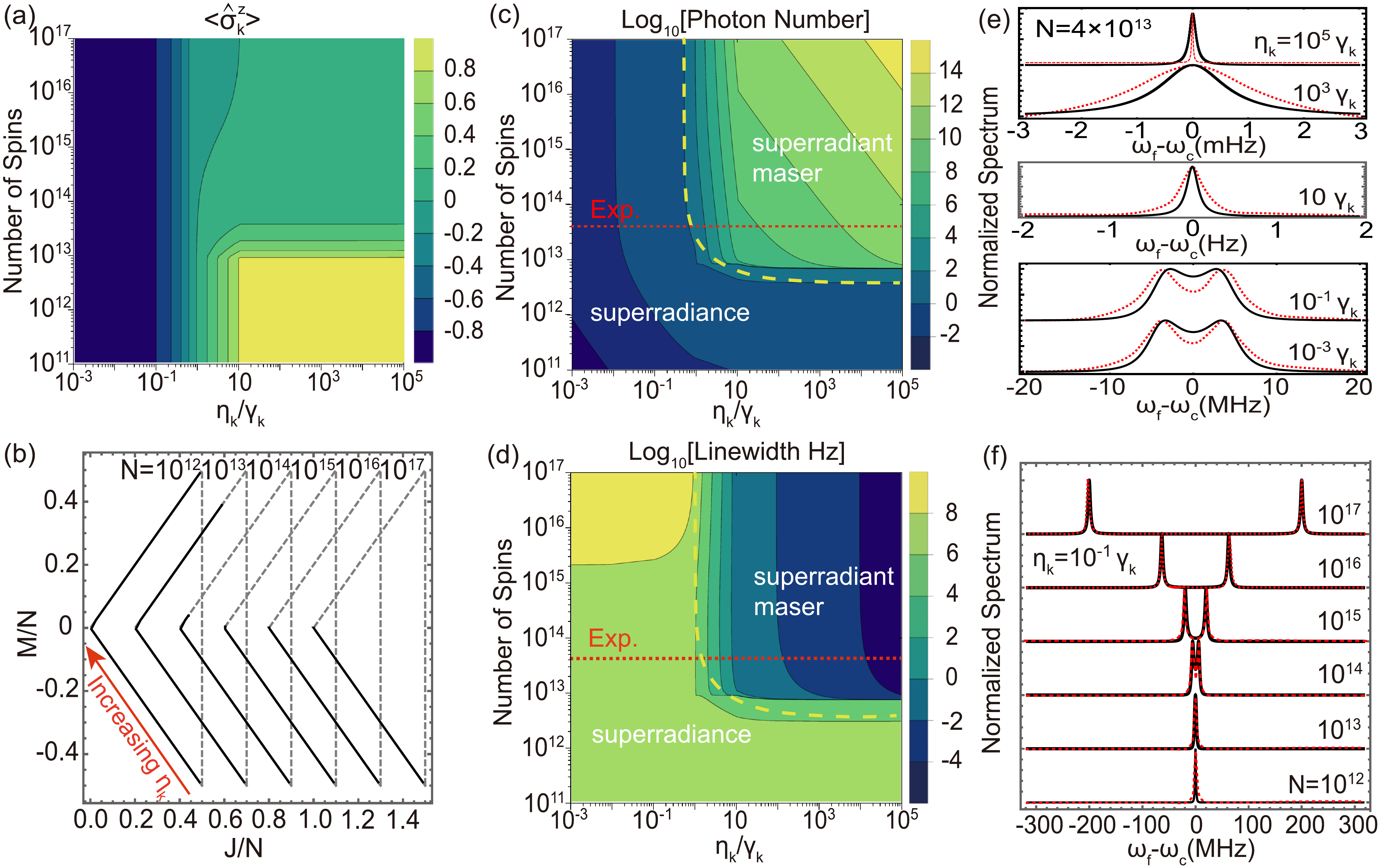}
\caption{\label{fig:number-spins} Influence of the number of $\mathrm{NV}^{-}$ spins $N$ on the evolution of the spin-ensemble (a,b) and the resonator (c-f) with increasing pumping rate $\eta_k$ (relative to the spin relaxation rate $\gamma_k$). Panel (a) shows the steady state population difference $\left \langle \hat{\sigma}_k^z \right \rangle$. Panel (b) shows the averaged Dicke state numbers $J,M$ normalized by $N$, bounded by the dashed gray lines. Panel (c) shows the intra-resonator photon number $\left \langle \hat{a}^\dagger \hat{a} \right \rangle$ and (d) shows the spectrum linewidth, where the yellow dashed curves indicate the transition boundaries calculated with Eq. \eqref{eq:boundaries}, and the red horizontal dashed lines indicate the number of spins $N=4\times 10^{13}$ as reported in the experiment \cite{JDBreeze}. Panel (e) and (f) show the normalized spectra (solid lines) with increasing  $\eta_k$  for fixed $N=4\times 10^{13}$, and with increasing $N$ for fixed $\eta_k=0.01\gamma_k$, respectively, where the red dashed lines are computed with Eq. \eqref{eq:comfrequencies}. In all the calculations, we assumed a dephasing rate $\chi_k=4$ MHz and a low temperature of $25$ mK. More details are given in the text. } 
\end{figure*}

As the calculations with the discretized inhomogeneous profile are too demanding to permit exhaustive studies of the dependence on various physical parameters and they offer also little analytical insight, we have recourse to a simpler description for those purposes. The comparison of the numerical results in Fig. \ref{fig:gaussian}(b) and Fig. \ref{fig:number-spins} (e), however, confirms the quantitative validity of these simpler calculations. Thus, in the following, we will rely on a simpler model to study how the superradiant maser is affected by the incoherent pumping rate, the number of $\mathrm{NV}^{-}$ spins, the temperature and the frequency detuning.


\subsection{Modelling of the Dynamics and Spectra of Identical Spins with a Large Homogeneous Broadening} \label{sec:HomBrodening}

With the ansatz that all spins are identical but subject to a large dephasing rate, the equations can be simplified dramatically as the correlations $\left\langle \hat{\sigma}_{k}^{\dagger}\hat{a}\right\rangle ,\left\langle \hat{\sigma}_{k}^{\dagger}\hat{b}\right\rangle $ and the population differences $\left\langle \hat{\sigma}_{k}^{z}\right\rangle $ are identical for any pseudo-spin $k$, and the correlations $\left\langle \hat{\sigma}_{k}^{\dagger}\hat{\sigma}_{k'}^{-}\right\rangle $ are identical for any pseudo-spin pair $\left(k,k'\right)$, see the Appendix E. This makes numerical calculations much easier.

The assumption of identical spins also permits a representation of the collective dynamics of the identical spins in the basis of Dicke states $\left|J,M\right\rangle $ \cite{RHDicke,LMandel}, where the half-integer or integer number $J\leq N/2$ reflects the permutation symmetry of the Dicke states, and the effective number of spins, which couple collectively with the resonator. We do not use the Dicke basis for numerical calculations, but the quantities calculated in the second-order mean-field approach do allow us to calculate the average of the numbers $M =\frac{1}{2}N \left\langle \hat{\sigma}_{k}^{z}\right\rangle $ and $J =\sqrt{\frac{3}{4}N +N \left(N -1\right)\left[\left\langle \hat{ \sigma}_{ k}^{\dagger}\hat{\sigma}_{ k'}^{-}\right\rangle +\frac{1}{4}\left\langle \hat{\sigma}_{ k}^{z}\right\rangle ^{2}\right]}$ from the values of the population difference $\left\langle \hat{\sigma}_{ k}^{z}\right\rangle $ and the spin-spin correlations $\left\langle \hat{ \sigma}_{ k}^{\dagger}\hat{\sigma}_{ k'}^{-}\right\rangle $ \cite{KDebnath}. 

The quantum numbers $J$ and $-J\leq M\leq J$ are useful indicators of the coupling and degree of excitation  of the spin ensemble \cite{RHDicke}. The energy levels of the Dicke states $\left|J,M\right\rangle $ for given $J$ form a ladder, and the ladders for different $J$ may be shifted horizontally to form a triangular pattern of the entire state space, see Fig.8 in the Appendix F. The dynamics of the system can be visualized as transitions among the Dicke states due to the spin-lattice relaxation, incoherent pumping, and dephasing, as well as the coherent coupling with the resonator, see the Appendix 
F. The Dicke state representation of our numerical solutions thus reveal the balance between these processes, and the transition to superradiance and superradiant masing as function of the different parameters.

\subsubsection{Influence of Number of $\mathrm{NV}^{-}$ Spins \label{sec:InfNumSpins}}

To reach the collective strong coupling regime and explore the superradiant maser, the collective coupling $\sqrt{N}g_{k}$ should be larger than any loss rate of the system. In Fig. \ref{fig:number-spins}, we investigate how the value of $N$ affects the response of the spin-ensemble (a,b) and the resonator (c-f) with increasing pumping rate $\eta_k$ (at low temperature $T=25$ mK). Here, we consider $N$ in the range of $10^{11}$ to $10^{17}$ and note that the case of $N\approx10^{13}$ and $10^{16}$ were studied in the experiments \cite{JDBreeze,AAngerer}. 

The excitation of the spin-ensemble is given by the population difference $\left \langle \hat{\sigma}_k^z \right \rangle$ [Fig. \ref{fig:number-spins} (a)] and the averaged Dicke state numbers $J,M$ [Fig. \ref{fig:number-spins} (b)]. Fig. \ref{fig:number-spins} (a) shows that the population difference is negative ($\left \langle \hat{\sigma}_k^z \right \rangle<0$) and positive ($\left \langle \hat{\sigma}_k^z \right \rangle>0$) for pumping rates smaller ($\eta_k<\gamma_k$) and larger than the spin relaxation rate ($\eta_k>\gamma_k$), respectively. $\left \langle \hat{\sigma}_k^z \right \rangle$ saturates for strong pumping, and 
it decreases from values close to unity for few spins $N<10^{13}$, to values near zero for many spins $N>6\times10^{13}$ (due to the balanced stimulated emission and absorption). Fig. \ref{fig:number-spins} (b) shows that with increasing pumping $\eta_k$ the spin ensemble evolves first along the lower rung of the Dicke ladders from the lower-right corner (i.e. the ground state) to the left-middle corner (sub-radiant states with low symmetry), and then along the upper rung of the Dicke ladders towards states close to the upper-right corner (i.e. fully excited state with high symmetry) \cite{KDebnath}.  Moreover, with increasing number of spins $N$, the quantum states reached by the spin ensemble at the strongest pumping rate, retracts from the upper-right corner towards the middle-left corner because the stimulated emission is balanced by the stimulated absorption leading to $M\approx0$ (zero population difference $\left\langle \hat{\sigma}_{k}^{z}\right\rangle \approx0$).  This evolution in the Dicke state basis is due to the interplay of quantum jumps associated with the individual spin relaxation, pumping, and dephasing and the coherent coupling with the resonator mode, see more information in the Appendix F.

The state of the resonator is represented by the mean photon number [Fig. \ref{fig:number-spins} (c)] and the spectrum linewidth [Fig. \ref{fig:number-spins} (d)] as well as the spectral features [Fig. \ref{fig:number-spins} (e,f)]. Fig. \ref{fig:number-spins} (c) shows that the photon number increases monotonously with increasing pumping $\eta_k$ when the number of $\mathrm{NV}^{-}$ spins $N$ is less than about $10^{13}$, while it increases dramatically for $\eta_k$ larger than the spin relaxation rate $\gamma_k$ when $N$ is much larger. The pumping threshold, over which the photon number increases dramatically, decreases and approaches $\gamma_k$ for large values of  $N$. The pumping threshold separates the superradiance regime with near or less than a single photon \cite{DMeiser} from the superradiant maser regime with many  photons, where stimulated photon emission and superradiance coexist and affect the radiation simultaneously \cite{DATieri,KDebnath,MANorcia}. 

To understand the transition between the different regimes, we have derived the condition for the presence of stimulated processes in  the superradiant maser regime, 
\begin{equation}
\frac{\eta_k}{\gamma_k} \geq \frac{ 2 n_k^{th}+1+N\mathcal{C}}{N\mathcal{C}-1} \label{eq:boundaries}
\end{equation}
with the single-particle  cooperativity $\mathcal{C} =   2k_{EET}/\kappa_c$, where $ k_{EET} = \frac{  2 g_k^2 (\lambda_k^s + \kappa/2)}{ ( \omega_{k} - \omega_{c})^2 + (\lambda_k^s + \kappa/2)^2 } $ is the energy transfer rate between the single spin and the resonator mode, see the Appendix G. For the resonant case ($\omega_c = \omega_k $, $\mathcal{C} =   4g_k^2/[(\lambda_k^s + \kappa_c/2)\kappa_c]$) and low temperature ($n_k^{th}\approx 0 $), as considered here,  the condition can be simplified as  $ \frac{\eta_k}{\gamma_k} \geq \frac{ 1+N\mathcal{C}}{N\mathcal{C}-1} $. The requirement $\eta_k >0$ leads to  $N \mathcal{C} > 1$ and thus $N > 1/\mathcal{C}$, which determines the lower-right part of the yellow dashed line in Fig. \ref{fig:number-spins} (c). For many spins, we have  $N \mathcal{C} \gg 1$ and thus $ \frac{\eta_k}{\gamma_k} \geq 1$, which determines the left part of the yellow dashed line.

Fig. \ref{fig:number-spins} (d) shows that the spectrum linewidth is several megahertz in the superradiance regime, but is below millihertz in the superradiant maser regime. These values are in agreement with the results of our numerical calculations with many sub-ensembles, as shown in Fig. \ref{fig:gaussian}(b). The large variation in linewidth is a consequence of the change of the spectrum from two peaks to a single peak, see Fig. \ref{fig:number-spins}(e,f). To interpret the spectral features, we have derived a semi-analytical expression for the complex frequencies of the peaks 
\begin{equation}
\tilde{\omega}_\pm = [\tilde{\omega}_k^* + \tilde{\omega}_c^* \pm \sqrt{(\tilde{\omega}_k^*-\tilde{\omega}_c^*)^2 - 4Ng_k^2  \left\langle \hat{\sigma}_{k}^{z}\right\rangle} ]/2, \label{eq:comfrequencies}
\end{equation}
where the term $N\left\langle \hat{\sigma}_{k}^{z}\right\rangle/2$ can be replaced by the average Dicke number $M$ or $\pm J$ according to Fig. \ref{fig:number-spins}(b), see the Appendix H. For the resonant condition $\omega_c =\omega_k$, as considered here, Eq. \eqref{eq:comfrequencies} can be simplified as $\tilde{\omega}_\pm = \omega_c + i (\lambda_k^s + \kappa_c/2 \pm \sqrt{R} )/2$ with $R = (\lambda_k^s -\kappa_c/2)^2 + 8g_k^2M$. For $R <0$,  $\tilde{\omega}_\pm = \omega_c\mp \sqrt{|R|}/2 +i (\lambda_k^s + \kappa_c/2)/2$, and we obtain two peaks centered around $\omega_c\mp \sqrt{|R|}/2$ (separated by $\sqrt{|R|}$) with the linewidths $\lambda_k^s + \kappa_c/2$, and if the peak separation is much less than the linewidth $\sqrt{|R|}\ll \lambda_k^s + \kappa_c/2$, the two peaks can not be discerned and they appear as a single peak. For $R >0 $, $\tilde{\omega}_\pm$ have the same real part $\omega_c $, but different imaginary parts $(\lambda_k^s + \kappa_c/2 \pm \sqrt{R} )/2$. As a result, the two peaks center around  $\omega_c $, and one peak has larger linewidth  $(\lambda_k^s + \kappa_c/2 +\sqrt{R} )/2$ and another one has smaller linewidth $(\lambda_k^s + \kappa_c/2 - \sqrt{R})/2$ (leading to the masing). 

For $N=4\times 10^{13}$ spins and increasing pumping rate $\eta_k$ as considered in Fig. \ref{fig:number-spins}(e), the Dicke state number $M$ increases gradually from large but negative value to zero, and then to small but positive value, which can be interpolated from the evolution of the Dicke state numbers for $N=10^{13}$ and $N=10^{14}$ in Fig. \ref{fig:number-spins} (b). As a result, the parameter $R$ is negative for small $\eta_k$ and positive for large $\eta_k$, and the spectrum changes from two broad peaks to a single narrow peak as shown in Fig. \ref{fig:number-spins}(e). In contrast, for the smaller pumping rate $\eta_k=0.1\gamma_k$ and increasing number of spins $N$, the population difference is negative and does not change while the Dicke state number $M$ and $J$ increase (because of increasing $N$). As a result, the parameter $R$ is always negative but its absolute value increases. In this case, we expect that the spectrum evolves from a single broad peak to two broad peaks, as shown in Fig. \ref{fig:number-spins}(f).

Since the spin-ensemble occupies always the lowest or upmost rung of Dick ladders ($J\approx |M|$) as shown in Fig. \ref{fig:number-spins}(b), we may apply the Holstein-Primakoff approximation \cite{JAGyamfi,THolstein} and describe the ensemble of identical spins for a given $J$ by bosonic creation $\hat{b}_J^\dagger$ and annihilation operators $\hat{b}_J$ of harmonic oscillators, leading to $\hat{H}_s^{\pm} \approx \hbar \omega_s \sum_J (\pm J \mp \hat{b}_J^\dagger \hat{b}_J)$, where the lower (upper) signs apply for weak pumping ($\hat{H}_s^{-}$) and strong pumping ($\hat{H}_s^{+}$), see the Appendix I. Thus, the spin-ensemble can be modeled as a normal and upside-down harmonic oscillator for the case with weak and strong pumping, respectively, see Fig. 9 in the Appendix I. It follows that we can approximate the spin-resonator interaction as the coupling between two harmonic oscillators, and obtain the beam-splitter-like coupling $H^-_{s-c} \approx \hbar \sum_J\sqrt{2J}g_s\left( \hat{b}_J^\dagger\hat{a}+\hat{a}^{\dagger}\hat{b}_J\right)$ and the parametric coupling $H^+_{s-c} \approx \hbar \sum_J\sqrt{2J}g_s\left( \hat{b}_J^\dagger\hat{a}^\dagger +\hat{a}^{\dagger}\hat{b}_J^\dagger \right)$ for the case with weak and strong pumping, respectively. In the low excitation limit, we can diagonalize the Hamiltonian $H^-_{s} + H^-_{s-c}$ and obtain the spin-photon dressed states, see Fig. A5, which lead to the double peaks in the spectrum in Fig. \ref{fig:number-spins}(e,f).  In contrast, the parametric coupling leads to an instability \cite{CKAndersen}, and is responsible for the masing action \cite{KDebnath}.

\begin{figure*}[htbp]
\centering \includegraphics[scale=0.3]{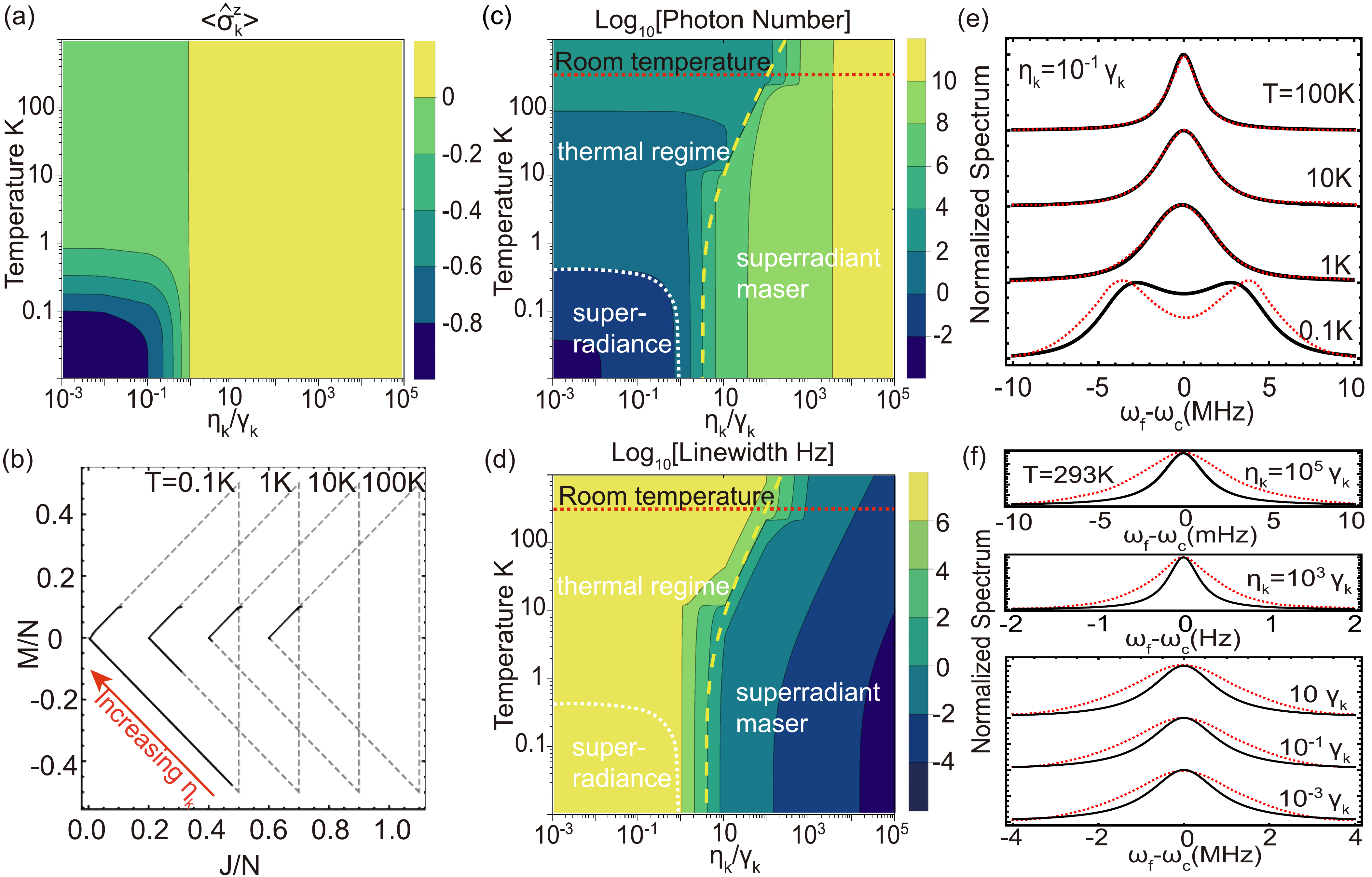} 
\caption{\label{fig:temperature} Influence of the temperature $T$  on the evolution of the spin ensemble (a,b) and the resonator (c-f)  with increasing pumping $\eta_k$. Panels (a) and (b) show the population difference $\left \langle \hat{\sigma}_k^z \right\rangle$ and the averaged Dicke state numbers $J,M$ (normalized to the number of spins $N$), respectively.  Panels (c) and (d) show the intra-resonator photon number $\left \langle \hat{a}^\dagger \hat{a} \right \rangle$ and the spectrum linewidth, respectively, where the regime with $\left \langle \hat{a}^\dagger \hat{a} \right \rangle>1$ for weak pumping is identified as the thermal regime,  the yellow dashed line is calculated with Eq. \eqref{eq:boundaries}, and the red horizontal dashed line marks the room temperature $T=293$ K. Panels (e) and (f) show the evolution of the normalized spectra (solid lines) with increasing temperature $T$ for fixed weak pumping $\eta_k=0.1\gamma_k$, and with increasing pumping rate $\eta_k$ for the system at room temperature $T=293$ K, respectively, where the dashed lines are computed with Eq. \eqref{eq:comfrequencies}. The number of $\mathrm{NV}^{-}$ spins is $N=4\times10^{13}$, as in the experiment \cite{JDBreeze}, and other parameters are the same as used in Fig. \ref{fig:number-spins}. }
\end{figure*}

\subsubsection{Influence of Temperature}

In contrast to lasing in the optical frequency domain, the maser operates in the  microwave frequency domain, and can be strongly affected by thermal excitation at energies $\sim k_{B}T$. Thus, we study  the influence of the temperature on  the spin state, the photon number and emission spectra for increasing pumping rate $\eta_k$ in Fig. \ref{fig:temperature}. Here, we consider  $N=4\times 10^{13}$ spins as reported in the experiment \cite{JDBreeze}.

The population difference $\left \langle \hat{\sigma}_k^z \right\rangle$  and the average Dicke state numbers $J,M$ normalized by the number of spins $N$ are shown in Fig. \ref{fig:temperature} (a,b). $\left \langle \hat{\sigma}_k^z \right\rangle$ approaches zero for weak pumping rates $\eta_k<\gamma_k$ at high temperature and for strong pumping $\eta_k>\gamma_k$ at any temperature due to the balanced thermal and stimulated processes, respectively. Fig. \ref{fig:temperature} (b) shows that the average numbers $J,M$ for  $T=0.1$ K are similar as for $0.025$ K [see Fig. \ref{fig:number-spins}(b)], and for no pumping, the absolute value of $J/N,M/N$ decreases to near zero with increasing temperature, which is consistent with the change of $\left \langle \hat{\sigma}_k^z \right\rangle$. 

Fig. \ref{fig:temperature} (c) shows that for high temperature and weak pumping, the photon number can become larger than unity, and for much higher temperature, the threshold pumping rate, above which the photon number increases dramatically, increases with increasing temperature (see the yellow dashed line). The threshold pumping rate can be estimated from Eq. \eqref{eq:boundaries}, which becomes $ \frac{\eta_k}{\gamma_k} \geq \frac{ 2n_k^{th}}{N\mathcal{C}} + 1 $ for the system with a large number of resonant spins (note  $N\mathcal{C}\gg 1$ and  $\mathcal{C} = 4g_k^2/[(\lambda_k^s + \kappa_c/2)\kappa_c]$ in this case). Thus, as the thermal population $n_k^{th}=[e^{\hbar\omega_k/(k_B T)} -1  ]^{-1}$ increases with increasing temperature $T$,  the threshold pumping rate increases.  In particular, at room temperature (red horizontal dashed line), the threshold pumping reaches about $160\gamma_k \approx 51$ Hz. The temperature, which separates the superradiant regime from the thermal regime, can be determined by requiring the thermal photon number equal to unity, i.e. $n_{c}^{th}=\left[e^{\hbar\omega_{c}/k_{B}T}-1\right]^{-1}=1$. 

Fig. \ref{fig:temperature} (d) shows that the spectrum linewidth remains in the megahertz range in the superradiance and thermal regime. In fact, the spectrum linewidth decreases a little with increasing temperature for weak pumping due to the change from two peaks to a single peak as shown in Fig. \ref{fig:temperature} (e). This spectral change can be understood from Eq. \eqref{eq:comfrequencies} or, more precisely, its resonant version $\tilde{\omega}_\pm = \omega_c\mp \sqrt{|R|}/2 +i (\lambda_k^s + \kappa_c/2)/2$ with $R = (\lambda_k^s -\kappa_c/2)^2 + 8g_k^2J$. Because the average number $J \approx M$ reduces to near zero with increasing temperature [Fig.\ref{fig:temperature} (b)],  the value of $R$ and thus $\sqrt{|R|}$  decreases, which leads to the final merging of the spectral peaks as observed here. Furthermore, for room temperature, we note in Fig. \ref{fig:temperature} (f) that the spectrum linewidth reduces from megahertz to millihertz when the pumping rate $\eta_k$ is higher than the threshold value (about $160\gamma_k \approx 51$ Hz).

\subsubsection{Influence of Frequency Detuning}

\begin{figure*}[htbp]
\centering \includegraphics[scale=0.3]{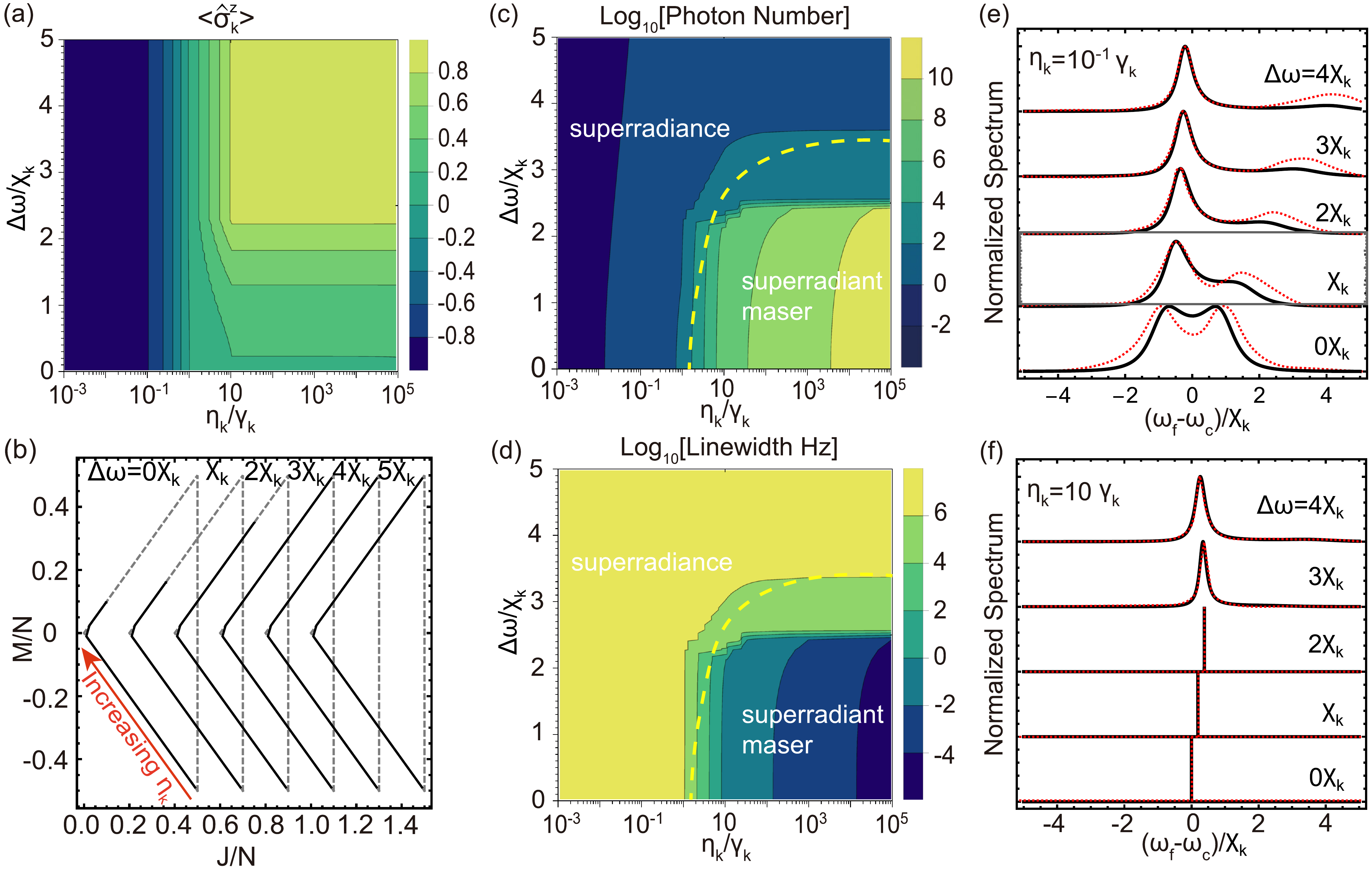}
\caption{\label{fig:detuning} Influence of frequency detuning  $\Delta\omega=\omega_{k}-\omega_{c}$ (in units of the dephasing rate $\chi_k$) between the  spin transition frequency $\omega_{k}$ and the resonator frequency $\omega_c$ on the state of the spins (a,b) and the resonator field (c-f) for different values of the pumping rate $\eta_k$. Panels (a) and (b) show the population difference $\left \langle \hat{\sigma}_k^z \right\rangle$ and the Dicke state numbers $J,M$ normalized to the number of spins $N$ (for $\Delta\omega=0$  to $5\chi_k$), respectively. Panels (c) and (d) show the intra-resonator photon number $\left \langle \hat{a}^\dagger \hat{a} \right \rangle$ and the spectral linewidth (in units of the dephasing rate $\chi_k$), respectively, where the yellow dashed line is computed with Eq. \eqref{eq:comfrequencies}. Panels (e) and (f) show  the spectra with increasing  $\Delta \omega$ from zero (lowest curve) to  $4\chi_k$ (topmost curve) for a weak pumping rate $\eta_k = 0.1\gamma_k$ and a strong pumping rate $\eta_k = 10 \gamma_k$, respectively. The dashed lines are computed with Eq. \eqref{eq:comfrequencies}. The number of $\mathrm{NV}^{-}$ spins is $N=4\times10^{13}$, as in the experiment \cite{JDBreeze}, the temperature is $T=25$ mK, and other parameters are the same as used in Fig. \ref{fig:number-spins}.}
\end{figure*}

In the optical lattice clock system \cite{MANorciaSciAdv,JGBohnet,MANorcia}, the frequency of the atomic transitions cannot be modified easily and thus careful design of the optical cavity is required to meet the resonance condition. In contrast, the transition frequency of the $\mathrm{NV}^{-}$ spins can be easily manipulated by adjusting the applied static magnetic field. This feature allows us to tune the spin ensemble into resonance with the microwave resonator. In addition, it also provides the possibility to actively control the performance of the superradiant maser. Thus, in Fig. \ref{fig:detuning}, we study the influence of the detuning $\Delta\omega=\omega_{k}-\omega_{c}$ between the  spin transition frequency $\omega_{k}$ and the resonator frequency $\omega_c$ on the response of the spin-ensemble and the resonator field to the increasing pumping rate $\eta_k$ for a low temperature ($T=25$ mK).

The spin state population difference $\left \langle \hat{\sigma}_k^z \right\rangle$ is shown in Fig. \ref{fig:detuning}(a), and the averaged Dicke state numbers $J,M$ normalized by the number of spins $N$ are shown in Fig. \ref{fig:detuning}(b). For pumping rates $\eta_k$ smaller than the relaxation rate $\gamma_k$, the negative $\left \langle \hat{\sigma}_k^z \right\rangle$ is independent of the frequency detuning, and the positive $\left \langle \hat{\sigma}_k^z \right\rangle$ for larger pumping rates $\eta_k>\gamma_k$ approaches zero for small frequency detuning $\Delta \omega <2\chi_k$ (due to the balanced stimulated emission and absorption), while it saturates to values close to unity for larger frequency detuning (due to the reduced stimulated processes). Fig. \ref{fig:detuning}(b) shows that the value of the normalized numbers $J/N,M/N$ for the largest pumping rate increases with increasing frequency detuning, which is consistent with the change of $\left \langle \hat{\sigma}_k^z \right\rangle$. 

Fig. \ref{fig:detuning} (c) shows that the intra-resonator photon number decreases by several orders of magnitude when the detuning
$\Delta\omega=\omega_{k}-\omega_{c}$ increases over several times of the dephasing rate $\chi_k$, and the threshold pumping rate, over which the superradiant masing occurs, increases with increasing $\Delta\omega$. The change of the pumping threshold can be understood by the simplified version of Eq. \eqref{eq:boundaries},  $\frac{\eta_k}{\gamma_k} \geq \frac{ 1+N\mathcal{C}}{N\mathcal{C}-1} $ for low temperature ($n_k^{th}\approx 0$) as considered here. Here, the single-particle  cooperativity $\mathcal{C} = 2k_{EET}/\kappa_c$ is given with the energy transfer rate $ k_{EET} = \frac{  2 g_k^2 (\lambda_k^s + \kappa/2)}{ (\Delta\omega)^2 + (\lambda_k^s + \kappa/2)^2} $.  As the frequency-detuning   $\Delta\omega$ increases, the energy transfer rate $k_{EET} $ and thus the cooperativity $\mathcal{C}$ decreases, which leads to the increased pumping threshold as verified by the yellow dashed line. 

Fig. \ref{fig:detuning} (d) shows that the spectrum linewidth does not change much with increasing frequency detuning  $\Delta\omega$ in the superradiance regime, but it decreases by several orders of magnitude in the superradiant maser regime.  A careful analysis shows that the spectrum in the former regime evolves from balanced double peaks to imbalanced double peaks, as shown in Fig. \ref{fig:detuning} (e), where the dominant peak near the resonator frequency is almost unchanged while the peak near the spin transition frequency becomes weaker. In the latter regime, the spectrum shows always a single peak, which shifts slightly  with increasing frequency detuning $\Delta\omega$, see Fig. \ref{fig:detuning} (f). This spectral change can be easily understood with Eq. \eqref{eq:comfrequencies}, see the dotted lines in Fig. \ref{fig:detuning} (e,f),  and the double peak feature can be attributed to the spin-photon dressed states, see Fig. 10 of the Appendix I. 

The analysis presented here permits an estimate of the effect on the emitted superradiant masing spectrum caused by changes of the spin and resonator frequencies due to magnetic field fluctuations and mechanical vibrations, respectively. If the system operates in the deep superradiant maser regime (with strong pumping $\eta_k$), the masing linewidth is not expected to fluctuate too much, as shown in Fig. \ref{fig:detuning} (d). From Fig. \ref{fig:detuning} (f), however, we estimate that the masing frequency $\omega_m$ changes by $\Delta \omega_m = 0.5\chi_k$ for the detuning $\Delta\omega = 2\chi_k$ leading to a pulling factor $\Delta\omega_m/\Delta\omega\approx 0.25$. Thus, the magnetic field and the resonator have to be stabilized so that the fluctuation of spin transition frequency and resonator is in the same order of magnitude as the desired spectral linewidth.

\section{Conclusion} 

In summary, we have studied  the possibility of achieving a  superradiant maser with $\mathrm{NV}^{-}$ center spins in diamond coupled to microwave resonators. Our study shows that a superradiant maser with a linewidth down to millihertz can be achieved if the spin-ensemble is pumped beyond a threshold rate, which can be as low as kilohertz for sufficiently many spins and a sufficiently low temperature. Our results are obtained by numerical solution of the second order mean field equations, and the threshold behavior of the system and the steady-state spectrum are well explained by analytical expressions. We have verified that the results are also obtained with a more elaborate theoretical modelling of the ensemble inhomogeneous broadening of the ensemble as the spin subensembles with different transition frequencies. This documents the robustness of the predicted superradiant masing. The superradiant masing with narrow linewidth should be readily achieved with the experimental setups currently available, and is also expected for systems with other solid spin dopants \cite{CBradac}, and might find applications in deep-space communications, radio astronomy, and high-precision metrology.

\begin{acknowledgments}
This work was supported by the National Natural Science Foundation of China through the project No. 12004344 and the project No. 62027816, as well as the Danish National Research Foundation through the Center of Excellence for Complex Quantum Systems (Grant agreement No. DNRF156).
\end{acknowledgments}

\appendix

\begin{figure}[htbp]
\begin{centering}
\includegraphics[scale=0.4]{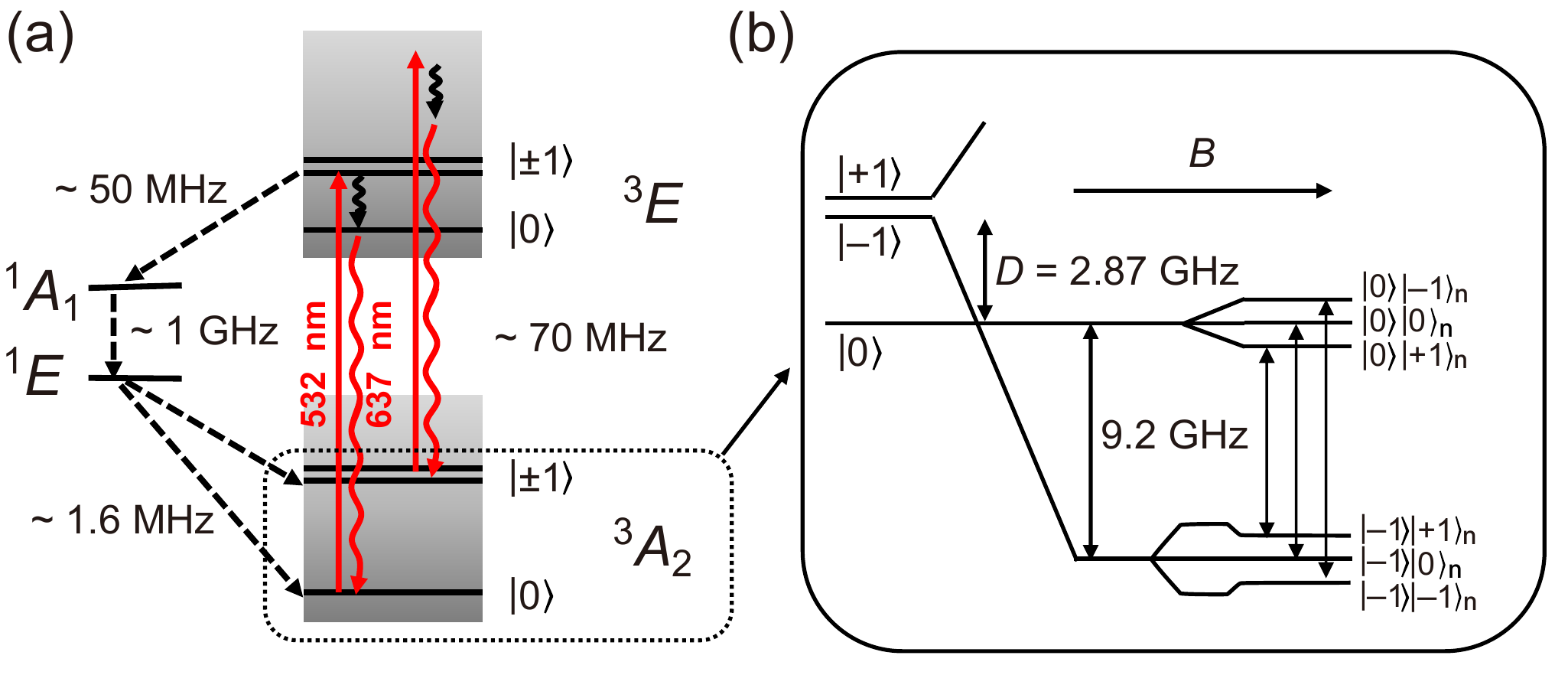}
\par\end{centering}
\caption{ \label{fig:spinlevels} Energy diagram with optical and microwave transitions in $\mathrm{NV}^{-}$ centers. Panel (a) shows the spin-triplet $^{3}A_{2}$, $^{3}E$ and spin-singlet $^{1}E$,$^{1}A_{1}$ electronic states and the transitions driven by optical pumping by a laser with wavelength $532$ nm, phononic relaxation (black arrows on gray background), fluorescence at the $637$ nm zero-phonon line, as well as non-radiative decay through the spin-singlet states (dashed lines). The rates of the various processes are listed next to the downward arrows, and the optical pumping and the decay process lead to an effective decay from the $\left|\pm 1\right\rangle $ state to the $\left|0 \right\rangle $ state. Panel (b) complements panel (a) with the Zeeman-shifted electron spin levels ($\left|+1\right\rangle $ upwards, and 
$\left|-1\right\rangle $ downwards), and the Zeeman-shifted nitrogen-nuclear spin levels $\left|\pm 1\right\rangle_n $ (the shifted direction depends on the electron spin levels) in the presence of a strong magnetic field, as well as the hyper-fine interaction between the electron and nuclear spin. Note that the electron and nuclear spin levels $\left|0\right\rangle$,  $\left|0\right\rangle_n$ are not affected by the magnetic field. Between the electron-nuclear spin states, there are three possible transitions, and we assume that only one of them is resonant to the resonator mode. See the text for more information.}
\end{figure}

\section{Energy Diagram of  Centers and Optical Pumping }{\label{app:spinlevels}}

In this appendix, we provide more information on the energy diagram of $\mathrm{NV}^{-}$ centers in diamond. We present the  pumping mechanism to create population inversion in the $\mathrm{NV}^{-}$ spin levels, and we demonstrate that the $\mathrm{NV}^{-}$ spins can be modeled as pseudo-1/2 spins for situation considered in the main text.  

The energy diagram of the $\mathrm{NV}^{-}$ centers is shown in Fig. \ref{fig:spinlevels}(a). The $\mathrm{NV}^{-}$ centers have one electronic ground state $^{3}A_{2}$ and three electronic excited states $^{3}E,^{1}A_{1},^{1}E$. The spin triplet states $^{3}A_{2},^{3}E$ are split into one state $\left|0\right\rangle $ and one doublet $\left|\pm1\right\rangle $ with the spin projection number $0$ and $\pm1$ along the quantization axes, respectively. The other states $^{1}A_{1},^{1}E$ are spin singlet states. The $\mathrm{NV}^{-}$ centers can be laser-excited  at 532 nm from the state $\left|0\right\rangle $ ($\left|\pm1\right\rangle $) of the $^{3}A_{2}$ state to the state $\left|0\right\rangle $ ($\left|\pm1\right\rangle $) of the $^{3}E$, which is often accompanied by the creation and subsequent relaxation of lattice phonons. The excited $\mathrm{NV}^{-}$ centers can return back to the state $\left|0\right\rangle $ ($\left|\pm1\right\rangle $) of the $^{3}A_{2}$ state by light emission at the $637$ nm zero-phonon line. In addition, the $\mathrm{NV}^{-}$ centers can decay non-radiatively from the  $\left|\pm1\right\rangle $ components of the $^{3}E$ state through the intermediate state $^{1}A_{1},^{1}E$ to the $\left|0\right\rangle $ and $\left|\pm1\right\rangle $ states of the ground state $^{3}A_{2}$. The optical excitation and the radiative emission do not change the spin projection, while the non-radiative process introduces an effective decay from the spin projection $\pm1$ to zero. Thus, by pumping the $\mathrm{NV}^{-}$ centers optically, we can create more population in the spin state $\left|0\right\rangle $ than in the other spin states $\left|\pm1\right\rangle $. In the presence of a strong magnetic field, the spin level $\left|-1\right\rangle $ can be shifted below the spin state $\left|0\right\rangle $, see Fig. \ref{fig:spinlevels} (b), and in this case, the optical pumping thus creates a population inversion between the upper $\left|0\right\rangle $ and lower $\left|-1\right\rangle $ spin level.

In the following, we give a detailed account of the energy shift of the electron spin states. We focus on the electronic ground state $^{3}A_{2}$, described by the Hamiltonian: 
\begin{equation}
\hat{H}_{e}=g_e \mu_{e}\mathbf{B}\cdot\hat{\mathbf{S}}+D\left[\hat{S}_{z}^{2}-\frac{1}{3}S\left(S+1\right)\right].\label{eq:electron-spin}
\end{equation}
Here, $g_{e}\approx2$ is the Lande g-factor and $\mu_{e}=9.274 \times 10^{-24} $ J/T is the Bohr magneton. $\mathbf{B}=\sum_{i=x,y,z}B_{i}\mathbf{e}_{i}$ and $\hat{\mathbf{S}}=\sum_{i=x,y,z}\hat{S}_{i}\mathbf{e}_{i}$ are the applied static magnetic field and the electron spin vector in the Cartesian coordinate system (with the unit vectors $\mathbf{e}_{i}$), respectively. In the basis of the spin states $\left\{ \left|+1\right\rangle ,\left|0\right\rangle ,\left|-1\right\rangle \right\}$, the $x$-, $y$- and $z$- component of the vector $\hat{\mathbf{S}}$ have the form 
\begin{align}
 & \hat{S}_{x}=\frac{1}{\sqrt{2}}\left(\begin{array}{ccc}
0 & 1 & 0\\
1 & 0 & 1\\
0 & 1 & 0
\end{array}\right),
\hat{S}{y}=\frac{1}{\sqrt{2}i}\left(\begin{array}{ccc}
0 & 1 & 0\\
-1 & 0 & 1\\
0 & -1 & 0
\end{array}\right), \nonumber 
\end{align}
\begin{align}
\hat{S}_{z}=\left(\begin{array}{ccc}
1 & 0 & 0\\
0 & 0 & 0\\
0 & 0 & -1
\end{array}\right).
\label{eq:SxSySz}
\end{align}
Using the above expressions, we can rewrite the second term of Eq. \eqref{eq:electron-spin} as
\begin{equation}
\frac{1}{3}D\left(\begin{array}{ccc}
1 & 0 & 0\\
0 & -2 & 0\\
0 & 0 & 1
\end{array}\right).
\end{equation}
This expression indicates that the state $\left|0\right\rangle $
has a frequency $-\frac{2}{3}D$ while the other states $\left|\pm1\right\rangle $
have the frequency $-\frac{1}{3}D$, leading to a transition frequency
$D=2\pi \times 2.87$ GHz in the absence of the magnetic field. The spin states $\left|\pm1\right\rangle $ are also split by local
electric fields due to strain in the diamond matrix. Since
this splitting is much smaller than the magnetic field-induced splitting as considered in the present article, we ignore it in the following discussion. However, we note that the former splitting might contribute also partially to the inhomogeneous broadening of the NV spin-ensemble. 

In the room-temperature maser based on $\mathrm{NV}^{-}$ spins \cite{JDBreeze}, the triplet nuclear spin of nitrogen atoms plays an important role. Thus, we need to complement the electron spin Hamiltonian given
by Eq. \eqref{eq:electron-spin} with the nuclear spin Hamiltonian
\begin{equation}
\hat{H}_{n}=-g_{n}\mu_{n}\mathbf{B}\cdot\hat{\mathbf{I}}+\hat{\mathbf{S}}\overleftrightarrow{A}\hat{\mathbf{I}}.\label{eq:nuclearspin}
\end{equation}
Here, $g_{n}$ is the nuclear Lande g-factor, $\mu_{n}=\mu_e/1837$ is the nuclear Bohr magneton, $\hat{\mathbf{I}}=\sum_{i=x,y,z}\hat{I}_{i}\mathbf{e}_{i}$
is the nuclear spin operator. The nucleus is a spin 1 particle and it has three states $\left|+1\right\rangle _{n},\left|0\right\rangle _{n},\left|-1\right\rangle _{n}$. In the basis of these states, the components $\hat{I}_{i}$ have same structure as  Eq. \eqref{eq:SxSySz}. The second term of Eq. \eqref{eq:nuclearspin} describes the hyper-fine interaction with the uni-axially anisotropic tensor $\overleftrightarrow{A}$, which
has the component $A_{\perp}=-2.7$ MHz and $A_{\parallel}=-2.1$ MHz with respect to the quantization axis. 

To describe the quantum states of the electron and nuclear spin together, in principle, we should introduce nine product states $\left|s_{e}\right\rangle \left|s_{n}\right\rangle _{n}$
(with $s_{e},s_{n}=0,\pm1$) and diagonalize the total spin Hamiltonian
$\hat{H}_{s}=\hat{H}_{e}+\hat{H}_{n}$. In view of large frequency difference between electron and nuclear momentum, it is, however, eligible to diagonalize first $\hat{H}_{e}$, and subsequently, for each electron spin eigenstate, diagonalize $\hat{H}_{n}$ among the nuclear spin states. 

In the following, we consider the particular case, where $\mathbf{B}$ is parallel to the line between the vacancy and the nitrogen atom, see Fig. \ref{fig:spinlevels} (b). In this case, Eq. \eqref{eq:electron-spin} can be easily diagonalized, and the energy of the spin state $\left|0\right\rangle $ is not changed, i.e. $\omega_{0}=-\frac{2}{3}D$,
while that of the spin states $\left|\pm1\right\rangle $ is given
by $\omega_{\pm}=\frac{1}{3}D\pm g_{e}\mu_{e}B$ (with the magnetic
field amplitude $B$). If the electron spin occupies the state $\left|0\right\rangle $,
we can ignore the hyper-fine interaction and simplify Eq. \eqref{eq:nuclearspin}
as $\hat{H}_{n}=-g_{n}\mu_{n}B \hat{I}_{z}$. As a result, the product state $\left|0\right\rangle \left|0\right\rangle _{n}$ has the frequency $\omega_{00}=\omega_{0}$ and the states $\left|0\right\rangle \left|\pm1\right\rangle _{n}$ have the frequencies $\omega_{0\pm1}=\omega_{0}\mp g_{n}\mu_{n}B$. If the electron spin occupies the state $\left|-1\right\rangle $,
Eq. \eqref{eq:nuclearspin} can be simplified as $-\left(g_{n}\mu_{n}B-A_{\parallel}\right)\hat{I}_{z}$.
As a result, the product state $\left|-1\right\rangle \left|0\right\rangle _{n}$ has the frequency $\omega_{-10}=\omega_{-}$ and the states $\left|-1\right\rangle \left|\pm1\right\rangle _{n}$
have the frequencies $\omega_{-1\pm1}=\omega_{-1}\pm\left(g_{n}\mu_{n}B-A_{\parallel}\right)$.
This analysis leads to the energy scheme shown in Fig. \ref{fig:spinlevels}
(b). Here, the presence of $A_{\parallel}$ makes the nuclear spin
state $\left|+1\right\rangle _{n}$ have the highest energy.
For the $\mathrm{NV}^{-}$ centers subject to an arbitrary magnetic field, we expect a similar energy scheme as Fig. \ref{fig:spinlevels} (b), however, with a more complex dependence of the energy levels on the magnetic field. 

When the analysis is extended to multiple $\mathrm{NV}^{-}$ centers, we have to account for that there are four possible orientations of the vacacny and the nitrogen atom. To maximize the Zeeman-shift for a given magnetic field, we can place the diamond in such way that the magnetic field is along one of these directions, see Fig. \ref{fig:spinlevels} (a). In this case, the magnetic field along the other quantization axes, and thus the splitting of the corresponding spin levels, are small. If the microwave resonator has a narrow linewidth it couples only to the spins with centers aligned along the magnetic field.Furthermore, for large magnetic field as considered in \citep{JDBreeze}, the transitions between different electron-nuclear spin states, as indicated in Fig. \ref{fig:spinlevels} (b), are well separated so that only one of them couples resonantly with the resonator mode. We assume that this simplification applies in this work so that we can treat  the resonant transition as the pseudo-1/2 spins.

\section{Second-order Mean-field Equations }{\label{sec:extra-equation}}

In the main text, we outline the procedure to derive the equations for the physical observables in the second-order mean-field approach, and explain the equations for the mean photon number in the main resonator $\left\langle \hat{a}^{\dagger}\hat{a}\right\rangle$ and the spin-photon correlation  $\left\langle \hat{\sigma}^{\dagger}_k \hat{a}\right \rangle$ and the mean photon number in the filter resonator $\left\langle \hat{b}^{\dagger}\hat{b}\right\rangle$, see Eq. (2), (3) and (4). In this appendix, we provide all the other equations needed for our calculations.  
The equation for the population difference $\left\langle \hat{\sigma}_{k}^{z}\right\rangle $ reads
\begin{align}
 & \frac{\partial}{\partial t}\left\langle \hat{\sigma}_{k}^{z}\right\rangle =-i2g_{k}\left(\left\langle \hat{\sigma}_{k}^{\dagger}\hat{a}\right\rangle -\left\langle \hat{a}^{\dagger}\hat{\sigma}_{k}^{-}\right\rangle \right) \nonumber \\
 & -\gamma_{k}\left[\left(2n_{k}^{th}+1\right)\left\langle \hat{\sigma}_{k}^{z}\right\rangle +1\right]-\eta_{k}\left(\left\langle \hat{\sigma}_{k}^{z}\right\rangle -1\right).\label{eq:inversion}
\end{align}
The equation for the spin-spin correlation $\left\langle \hat{\sigma}_{k}^{\dagger}\hat{\sigma}_{k'}^{-}\right\rangle $
reads 
\begin{align}
& \frac{\partial}{\partial t}\left\langle \hat{\sigma}_{k}^{\dagger}\hat{\sigma}_{k'}^{-}\right\rangle =i\left(\tilde{\omega}_{k}^{*}-\tilde{\omega}_{k'}\right)\left\langle \hat{\sigma}_{k}^{\dagger}\hat{\sigma}_{k'}^{-}\right\rangle \nonumber \\
& +i\left(g_{k'}\left\langle \hat{\sigma}_{k}^{\dagger}\hat{a}\right\rangle \left\langle \sigma_{k'}^{z}\right\rangle -g_{k}\left\langle \hat{\sigma}_{k}^{z}\right\rangle \left\langle a^{\dagger}\hat{\sigma}_{k'}^{-}\right\rangle \right).\label{eq:spinspincorrelation}
\end{align}

The equation for the photon-photon correlation $\left\langle \hat{a}^{\dagger}\hat{b}\right\rangle $
reads 
\begin{align}
 & \frac{\partial}{\partial t}\left\langle \hat{a}^{\dagger}\hat{b}\right\rangle =i\left(\tilde{\omega}_{c}^{*}-\tilde{\omega}_{f}\right)\left\langle \hat{a}^{\dagger}\hat{b}\right\rangle \nonumber \\
 & +i\sum_{k}g_{k}\left\langle \hat{\sigma}_{k}^{\dagger}\hat{b}\right\rangle +iG\left(\left\langle \hat{b}^{\dagger}\hat{b}\right\rangle -\left\langle \hat{a}^{\dagger}\hat{a}\right\rangle \right), 
\end{align}
which depends on the correlations between the spins and the filter
resonator $\left\langle \hat{\sigma}_{k}^{\dagger}\hat{b}\right\rangle $ (and the conjugation
$\left\langle \hat{b}^{\dagger} \hat{\sigma}_{k}\right\rangle $). The equations for these correlations read
\begin{align}
 & \frac{\partial}{\partial t}\left\langle \hat{\sigma}_{k}^{\dagger}\hat{b}\right\rangle =i\left(\tilde{\omega}_{k}^{*}-\tilde{\omega}_{f}\right)\left\langle \hat{\sigma}_{k}^{\dagger}\hat{b}\right\rangle  -ig_{k}\left\langle \hat{a}^{\dagger}\hat{b}\right\rangle \left\langle \hat{\sigma}_{k}^{z}\right\rangle -iG\left\langle \hat{\sigma}_{k}^{\dagger}\hat{a}\right\rangle .\label{eq:spinfilterphoton}
\end{align}

\begin{table*}[t]
\begin{centering}
\resizebox{\textwidth}{30mm}{
\begin{tabular}{c|c|c|c|c|c|c}
\hline 
Ref. & \citenum{JDBreeze}$^*$ & \citenum{AAngerer} &  \citenum{SPutz} & \citenum{RAmsuss} & \citenum{YKubo} & \citenum{AAngerer1} \tabularnewline
\hline 
$\omega_{c}/2\pi$ & $9.22$ GHz & $3.18$ GHz & $2.69$ GHz  & $2.90$ GHz  & $2.87$ GHz   &  $3.12$ GHz  \tabularnewline
\hline 
$\kappa_{c}/2\pi$ & $0.3$ MHz & $13.8$ MHz   & $0.8$ MHz  & $0.8$ MHz  & $1.5$ MHz   & $3.82$ MHz  \tabularnewline
\hline 
$N$ & $4\times10^{13}$ & $1.5\times10^{16}$ & $2.5\times10^{12}$ & $10^{12}$ & $10^{12}$ & $10^{17}$\tabularnewline
\hline 
 $(\chi_{k}/2\pi)^{**}$ & $0.64$ MHz & $4.7$ MHz   & $2.6$ MHz  & - & $3$ MHz  & $3$ MHz   \tabularnewline
\hline 
$g_{k}/2\pi$ & $0.11$ Hz & $0.051$ Hz & $12$ Hz  & $12$ Hz   & $12$ Hz  & 0.07 Hz \tabularnewline
\hline 
$\Gamma_c/2\pi$ & $1.04\times 10^{-6}$ Hz & $7.5\times10^{-10}$ Hz  & $7.2\times 10^{-4}$ Hz & $7.2\times 10^{-4}$ Hz & $3.84\times 10^{-4}$ Hz & $5.13\times 10^{-9}$ Hz\tabularnewline
\hline 
$\Omega/2\pi$ & $0.70$ MHz & $6.12$ MHz & $19$ MHz & $12$ MHz & $12$ MHz  & $12$ MHz \tabularnewline
\hline 
Regimes & Strong & Weak & Strong & Strong$^{***}$ & Strong & Strong \tabularnewline
\hline 
\end{tabular}}
\par\end{centering}
\begin{centering}
\caption{\label{tab:params} Summary of experimental parameters for the microwave resonators (frequency $\omega_c$, damping rate $\kappa_c$), the $\mathrm{NV}^{-}$ center spins (number of spins $N$, dephasing rate $\chi_k$, single spin-resonator coupling $g_k$), as well as the combined parameters (Purcell-enhanced decay rate $\Gamma_c$, collective spin-resonator coupling $\Omega=\sqrt{N}g_k$). The last line assigns the systems to either the  regime of collective weak or strong coupling.  $*$ marks the parameters (in the first column) used in our numerical simulations. $**$ indicates that the dephasing rate corresponds to half of the FWHM of the inhomogeneously broadened spin spectrum. $***$ marks that a dephasing rate was not reported in the experiment, and the strong coupling regime is deduced by using the typical value of the dephasing rate.}.
\end{centering}
\end{table*}

\section{Summary of Experimental Parameters }{\label{app:parameters}}

In Tab. \ref{tab:params}, we summarize parameters reported in several experiments on the microwave resonators (frequency $\omega_c$, damping rate $\kappa_c$) and the $\mathrm{NV}^{-}$ center spins (number of spins $N$, dephasing rate $\chi_k$, single spin-resonator coupling $g_k$), as well as the combined parameters (the collective Rabi-frequency $\Omega = \sqrt{N}g_k$ and the Purcell-enhanced decay rate $\Gamma_c = 4g_k^2/\kappa_c$). If the parameter $\Omega$ is larger or smaller than $\kappa_c$ and $\chi_k$, the system works in the collective strong or weak coupling regime (see the the last line of Tab. \ref{tab:params}). Note that the spin relaxation rate $\gamma_k$ is usually orders of magnitude smaller than the dephasing rate $\chi_k$ and is thus usually not reported in the experiments. We can utilize the parameter $\Gamma_c$ to estimate the orders of magnitude for the linewidth in the superradiant maser regime.

\section{Second-order Mean-field Equations for Spin Sub-ensembles and Related Results}{\label{sec:groupsequations}}

In Sec. IV.A of the main text, we presented the emission spectra for the system of spin sub-ensembles representing the inhomogeneous broadening of the spins. To this end, we adopted the second-order mean-field equations given in Sec. III of in the main text to the system with spin sub-ensembles labeled by $\alpha$. To reduce the computational effort, we assume that the spins in each sub-ensemble are identical, and obtain the following equations for the quantities related to individual spin sub-ensembles and between spin sub-ensembles.

\subsection{Equations Involving Main Resonator}

The mean intra-resonator photon number follows the equation 
\begin{align}
& \frac{\partial}{\partial t}\left\langle \hat{a}^{\dagger}\hat{a}\right\rangle =-\kappa_{c}\left\langle \hat{a}^{\dagger}\hat{a}\right\rangle +\kappa_{c}n_{c}^{th} \nonumber \\
&+i\sum_{\alpha}N_{\alpha}g_{\alpha}\left(\left\langle \hat{\sigma}_{\alpha k}^{\dagger}\hat{a}\right\rangle -\left\langle \hat{a}^{\dagger}\hat{\sigma}_{\alpha k}^{-}\right\rangle \right),\label{eq:apa_alpha}
\end{align}
where $N_{\alpha},g_{\alpha}$ denote the number of spins and the spin-photon coupling of the $\alpha$-th spin sub-ensemble, respectively. The spin-photon correlation $\left\langle \hat{\sigma}_{\alpha k}^{\dagger}\hat{a}\right\rangle $ follows the equation 
\begin{align}
 & \frac{\partial}{\partial t}\left\langle \hat{\sigma}_{\alpha k}^{\dagger}\hat{a}\right\rangle =i(\tilde{\omega}_{\alpha k}^* -\tilde{\omega}_c)\left\langle \hat{\sigma}_{\alpha k}^{\dagger}\hat{a}\right\rangle \nonumber \\
 & -ig_{\alpha k}\left\langle \hat{a}^{\dagger}\hat{a}\right\rangle \left\langle \hat{\sigma}_{\alpha k}^{z}\right\rangle -i(g_{\alpha k}/2)\left(\left\langle \hat{\sigma}_{\alpha k}^{z}\right\rangle +1\right) \nonumber \\
 & +ig_{\alpha k}\left\langle \hat{\sigma}_{\alpha k}^{\dagger} \hat{\sigma}_{\alpha k'}^{-}\right\rangle -i\sum_{\alpha'}N_{\alpha'}g_{\alpha'}\left\langle \hat{\sigma}_{\alpha k}^{\dagger}\hat{\sigma}_{\alpha'k'}^{-}\right\rangle .\label{eq:sigmapa}
\end{align}
Here, we have defined $\tilde{\omega}_{\alpha k}=\omega_{\alpha k}-i\lambda_{\alpha k}^{s}$ with the total dephasing rate of pseudo-spins $\lambda_{\alpha k}^{s}=\frac{1}{2}\left[\gamma_{\alpha k}\left(2n_{\alpha k}^{th}+1\right)+\eta_{\alpha k}\right]+\chi_{\alpha k}$. In the last line of the above equation, the first and second term
describe the spin-spin correlation in individual spin sub-ensemble and between the spin sub-ensembles, respectively. The population difference $\left\langle \hat{\sigma}_{\alpha k}^{z}\right\rangle $
and the spin-spin correlation $\left\langle \hat{\sigma}_{\alpha k}^{\dagger}\hat{\sigma}_{\alpha'k'}^{-}\right\rangle $
follow the equations 
\begin{align}
 & \frac{\partial}{\partial t}\left\langle \hat{\sigma}_{\alpha k}^{z}\right\rangle =-i2g_{\alpha}\left(\left\langle \hat{\sigma}_{\alpha k}^{\dagger}\hat{a}\right\rangle -\left\langle \hat{a}^{\dagger}\hat{\sigma}_{\alpha k}^{-}\right\rangle \right) \nonumber \\
 &-\gamma_{\alpha k}\left[\left(2n_{\alpha k}^{th}+1\right)\left\langle \hat{\sigma}_{\alpha k}^{z}\right\rangle +1\right]-\eta_{\alpha k}\left(\left\langle \hat{\sigma}_{\alpha k}^{z}\right\rangle -1\right),\label{eq:sigmaz} \\
 & \frac{\partial}{\partial t}\left\langle \hat{\sigma}_{\alpha k}^{\dagger}\hat{\sigma}_{\alpha'k'}^{-}\right\rangle =i\tilde{\omega}_{\alpha k\alpha'k'}\left\langle \hat{\sigma}_{\alpha k}^{\dagger}\hat{\sigma}_{\alpha'k'}^{-}\right\rangle  \nonumber \\
 & +i\left(g_{\alpha'}\left\langle \hat{\sigma}_{\alpha k}^{\dagger}\hat{a}\right\rangle \left\langle \hat{\sigma}_{\alpha'k'}^{z}\right\rangle -g_{\alpha}\left\langle \hat{\sigma}_{\alpha k}^{z}\right\rangle \left\langle \hat{a}^{\dagger}\hat{\sigma}_{\alpha'k'}^{-}\right\rangle \right).\label{eq:sigmapsigmam}
\end{align}
Here, the complex frequency is defined as  $\tilde{\omega}_{\alpha k\alpha'k'} = \tilde{\omega}_{\alpha k}^*  - \tilde{\omega}_{\alpha'k'}$.

To reduce the computational effort, we consider the steady-state version of the equations in the previous paragraph:
\begin{align}
&\left\langle \hat{a}^{\dagger}\hat{a}\right\rangle =i\sum_{\alpha}N_{\alpha}\frac{g_{\alpha k}}{\kappa_{c}}\left(\left\langle \hat{\sigma}_{\alpha k}^{\dagger}\hat{a}\right\rangle -\left\langle \hat{a}^{\dagger}\hat{\sigma}_{\alpha k}^{-}\right\rangle \right)+n_{c}^{th},\label{eq:sphoton} \\
&\left\langle \hat{\sigma}_{\alpha k}^{z}\right\rangle =\frac{-i2g_{\alpha}\left(\left\langle \hat{\sigma}_{\alpha k}^{\dagger}\hat{a}\right\rangle -\left\langle \hat{a}^{\dagger}\hat{\sigma}_{\alpha k}^{-}\right\rangle \right)+\eta_{\alpha k}-\gamma_{\alpha k}}{\gamma_{\alpha k}\left(2n_{\alpha k}^{th}+1\right)+\eta_{\alpha k}},\label{eq:sinversion} \\
&\left\langle \hat{\sigma}_{\alpha k}^{\dagger}\hat{\sigma}_{\alpha'k'}^{-}\right\rangle =\tilde{\omega}_{\alpha k\alpha'k'}^{-1}(g_{\alpha k}\left\langle \hat{\sigma}_{\alpha k}^{z}\right\rangle  \left\langle \hat{a}^{\dagger}\hat{\sigma}_{\alpha'k'}^{-}\right\rangle \nonumber \\
&-g_{\alpha'}\left\langle \hat{\sigma}_{\alpha k}^{\dagger}\hat{a}\right\rangle \left\langle \hat{\sigma}_{\alpha'k'}^{z}\right\rangle),\label{eq:sscorrelation}
\end{align}
as well as 
\begin{align}
 & i(\tilde{\omega}_{\alpha k}^* - \tilde{\omega}_c) \left\langle \hat{\sigma}_{\alpha k}^{\dagger}\hat{a}\right\rangle =ig_{\alpha}\left(\left\langle \hat{a}^{\dagger}\hat{a}\right\rangle \left\langle \hat{\sigma}_{\alpha k}^{z}\right\rangle -\left\langle \hat{\sigma}_{\alpha k}^{\dagger} \hat{\sigma}_{\alpha k'}^{-}\right\rangle \right)\nonumber \\
 & +i\sum_{\alpha'}N_{\alpha'}g_{\alpha'}\left\langle \hat{\sigma}_{\alpha k}^{\dagger}\hat{\sigma}_{\alpha'k'}^{-}\right\rangle +i\frac{g_{\alpha}}{2}\left(\left\langle \hat{\sigma}_{\alpha k}^{z}\right\rangle +1\right).\label{eq:scorrelation}
\end{align}
Inserting  Eq. \eqref{eq:sphoton}, Eq. \eqref{eq:sinversion} and Eq. \eqref{eq:sscorrelation} into  Eq. \eqref{eq:scorrelation}, we can obtain the self-consistent equations for the spin-photon correlations $ \left\langle \hat{\sigma}_{\alpha k}^{\dagger}\hat{a}\right\rangle$ and $ \left\langle \hat{a}^\dagger \hat{\sigma}_{\alpha k}\right\rangle$. Then, using the FindRoot program in the Mathematica, we can easily solve the coupled equations for these correlations for systems with many spin sub-ensembles. 

\subsection{Equations Involving Filter Resonator}

To compute the steady-state spectrum, the photon number $\left\langle \hat{b}^{\dagger}\hat{b}\right\rangle $ in the filter resonator do not change, and the photon-photon correlation $\left\langle \hat{a}^{\dagger}\hat{b}\right\rangle $ and the spin-photon correlation $\left\langle \hat{\sigma}_{\alpha k}^{\dagger}\hat{b}\right\rangle $ follow
the equations 
\begin{align}
 & \frac{\partial}{\partial t}\left\langle \hat{a}^{\dagger}\hat{b}\right\rangle =i\tilde{\omega}_{cf}\left\langle \hat{a}^{\dagger}\hat{b}\right\rangle   +i\sum_{\alpha}N_{\alpha}g_{\alpha k}\left\langle \hat{\sigma}_{\alpha k}^{\dagger}\hat{b}\right\rangle \nonumber 
 \\
 & + iG\left(\left\langle \hat{b}^{\dagger}\hat{b}\right\rangle 
 -\left\langle \hat{a}^{\dagger}\hat{a}\right\rangle \right),\label{eq:apb} \\
 & \frac{\partial}{\partial t}\left\langle \hat{\sigma}_{\alpha k}^{\dagger}\hat{b}\right\rangle =i\tilde{\omega}_{\alpha kf}\left\langle \hat{\sigma}_{\alpha k}^{\dagger}\hat{b}\right\rangle  -ig_{\alpha}\left\langle \hat{a}^{\dagger}\hat{b}\right\rangle \left\langle \hat{\sigma}_{\alpha k}^{z}\right\rangle -iG\left\langle \hat{\sigma}_{\alpha k}^{\dagger}\hat{a}\right\rangle .\label{eq:siamgpb}
\end{align}
Here, we have introduced the complex frequency $\tilde{\omega}_{cf}=\omega_{c}-\omega_{f}+i\frac{1}{2}\left(\kappa_{c}+\kappa_{f}\right)$ and $\tilde{\omega}_{\alpha k f}=\omega_{\alpha k }-\omega_{f}+i\frac{1}{2}\left(2 \lambda_{\alpha k }^s+\kappa_{f}\right)$. 

To reduce the computational effort, we consider the steady-state solution of
Eq. \eqref{eq:siamgpb} :
\begin{equation}
\left\langle \hat{\sigma}_{\alpha k}^{\dagger}\hat{b}\right\rangle =\tilde{\omega}_{\alpha kf}^{-1}\left(g_{\alpha}\left\langle \hat{a}^{\dagger}\hat{b}\right\rangle \left\langle \hat{\sigma}_{\alpha k}^{z}\right\rangle +G\left\langle \hat{\sigma}_{\alpha k}^{\dagger}\hat{a}\right\rangle \right).
\end{equation}
Inserting the above expression to Eq. \eqref{eq:apb} , we get
the steady-state photon-photon correlation
\begin{equation}
\left\langle \hat{a}^{\dagger}\hat{b}\right\rangle =G\frac{\left\langle \hat{a}^{\dagger}\hat{a}\right\rangle -\left\langle \hat{b}^{\dagger}\hat{b}\right\rangle -\sum_{\alpha}\tilde{\omega}_{\alpha kf}^{-1}N_{\alpha}g_{\alpha}\left\langle \hat{\sigma}_{\alpha k}^{\dagger}\hat{a}\right\rangle }{\tilde{\omega}_{cf}+\sum_{\alpha}\tilde{\omega}_{\alpha kf}^{-1}N_{\alpha}g_{\alpha}^{2}\left\langle \hat{\sigma}_{\alpha k}^{z}\right\rangle }.
\end{equation}
Inserting the above results to  Eq.(4) in the main text, we get the
steady-state mean photon number in the filter cavity 
\begin{equation}
\left\langle \hat{b}^{\dagger}\hat{b}\right\rangle =\frac{G^{2}2\mathrm{Im}\left[\frac{\left\langle \hat{a}^{\dagger}\hat{a}\right\rangle -\sum_{\alpha}\tilde{\omega}_{\alpha kf}^{-1}N_{\alpha}g_{\alpha}\left\langle \hat{\sigma}_{\alpha k}^{\dagger}\hat{a}\right\rangle }{\tilde{\omega}_{cf}+\sum_{\alpha}\tilde{\omega}_{\alpha kf}^{-1}N_{\alpha}g_{\alpha}^{2}\left\langle \hat{\sigma}_{\alpha k}^{z}\right\rangle }\right]}{G^{2}2\mathrm{Im}\left[\tilde{\omega}_{cf}+\sum_{\alpha}\omega_{\alpha kf}^{-1}N_{\alpha}g_{\alpha}^{2}\left\langle \hat{\sigma}_{\alpha k}^{z}\right\rangle \right]^{-1}-\kappa_{f}}.\label{eq:bpb-semi-analytical}
\end{equation}

\subsection{Syncronization of Spin Subensembles} \label{sec:Syncronization}

\begin{figure}
\centering \includegraphics[scale=0.32]{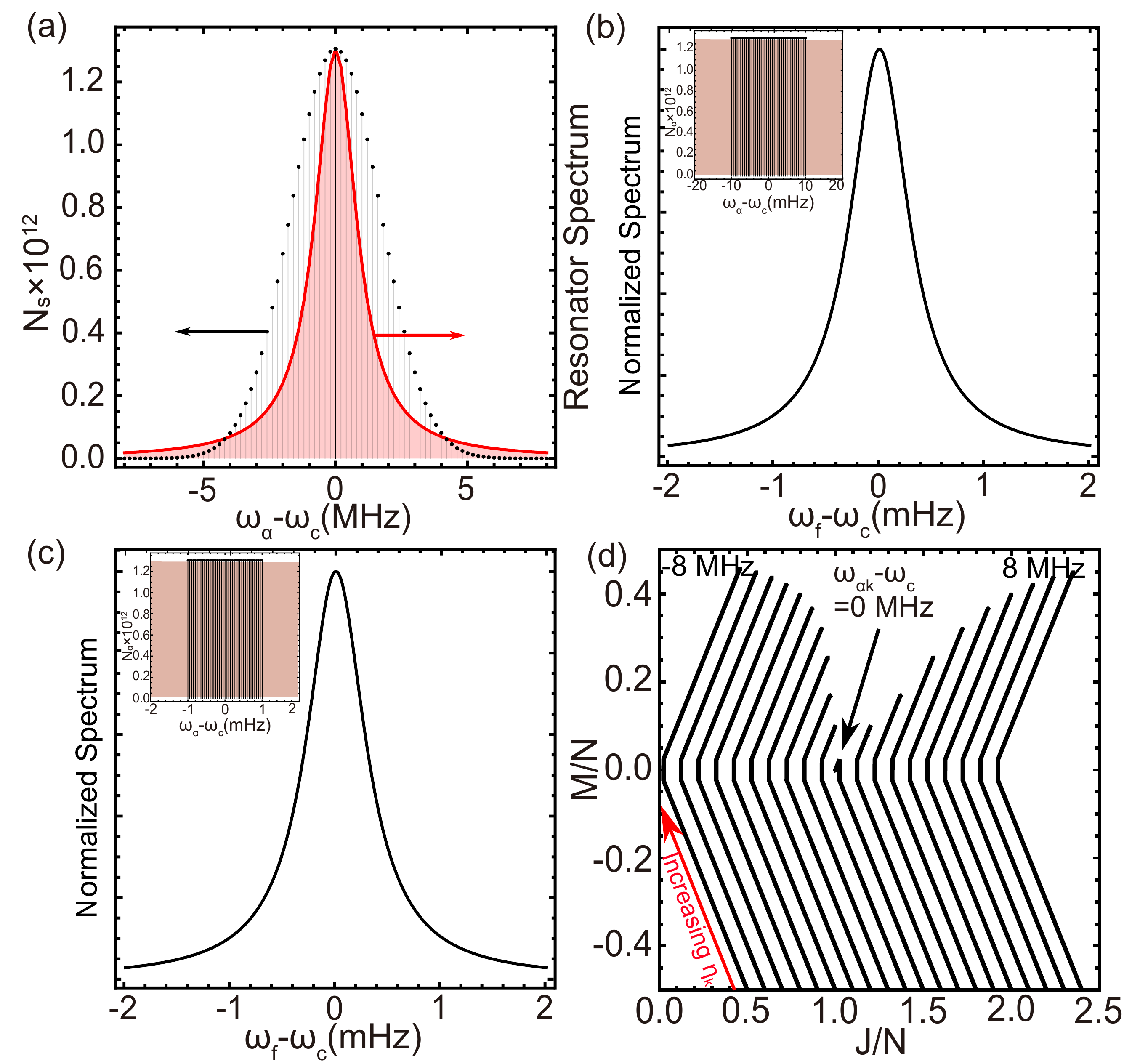}
\caption{\label{fig:Dickesub-ensembles} Supplemental results for spin subensembles. Panel (a) is similar to Fig. 2(a) in the main text except that the single spin subensemble, which is at the center of Gaussian distribution and is ideally resonant to the resonator, splits into many subsubensembles. Panel (b) and (c) show that the spectrum with linewidth in the millihertz range does not change if the center subsubensembles spread about $20$ mHz and $2$ mHz wide (insets), respectively. Here, the total number of spins is $8\times 10^{13}$ and the spin pumping rate is $\eta_k = 10^3\gamma_k$.  Panel (d) shows the evolution of the Dicke states of the individual spin sub-ensembles with increasing pumping $\eta_k$, where the numbers mark the frequency detuning of the spin sub-ensemble with respect to the resonator. For visualization purposes, we have considered $20$ spin sub-ensembles here. The temperature is $25$ mK, and other parameters are specified in the main text. }
\end{figure}

In Fig. 2(b) of the main text, we observed a sharp emission peak with linewidth in the millihertz range for the system with $50$ spin sub-ensembles when the incoherent pumping rate $\eta_k$ exceeds the spin relaxation rate $\gamma_k$. Fig. \ref{fig:Dickesub-ensembles} (a-c) show that this sharp emission peak persists even if we further discretize the center spin subensemble into many subsubensembles (see the insets). Here, we attribute the insensitivity of the spectrum  to the synchronization effects of the sub-ensembles \citep{AShankar,MXu,KDebnath1}. To support this, we look at the evolution of the Dicke states of each  spin sub-ensembles with increasing $\eta_k$, see Fig. \ref{fig:Dickesub-ensembles}(d). We see that the spin sub-ensembles nearly resonant to the resonator explore the Dicke states with low symmetry due to the balanced stimulated emission and absorption, while the spin sub-ensembles off-resonant to the resonator explore the Dicke states of higher symmetry close to the upper-right corner where the coupling with the resonator is reduced. These results suggest that all the spins within the inhomogeneous broadening are excited and contribute to the spectrum, while the spins closer to resonance with the resonator contribute more than the other sub-ensembles.

\section{System with Identical Spins}{\label{sec:identicalspins}}

In the following, we present the simplified equations for the systems
with identical spins. In this case, the mean photon number $\left\langle \hat{a}^{\dagger}\hat{a}\right\rangle $ follows the equation 
\begin{align}
 & \frac{\partial}{\partial t}\left\langle \hat{a}^{\dagger}\hat{a}\right\rangle =-\kappa_c\left\langle \hat{a}^{\dagger}\hat{a}\right\rangle +\kappa_c n_{c}^{th} +iNg_{k}\left(\left\langle \hat{\sigma}_{k}^{\dagger}\hat{a}\right\rangle -\left\langle a^{\dagger}\hat{\sigma}_{k}^{-}\right\rangle \right), \label{eq:apa_ide}
\end{align}
while the spin-photon correlation follows the equation 
\begin{align}
 & \frac{\partial}{\partial t}\left\langle \hat{\sigma}_{k}^{\dagger}\hat{a}\right\rangle =i\left(\tilde{\omega}_{k}^{*}-\tilde{\omega}_{c}\right)\left\langle \hat{\sigma}_{k}^{\dagger}\hat{a}\right\rangle -ig_{k}\left\langle \hat{a}^{\dagger}\hat{a}\right\rangle \left\langle \hat{\sigma}_{k}^{z}\right\rangle  \nonumber \\
 & -ig_{k}\frac{1}{2}\left(\left\langle \hat{\sigma}_{k}^{z}\right\rangle +1\right)-i\left(N-1\right)g_{k}\left\langle \hat{\sigma}_{k}^{\dagger}\hat{\sigma}_{k'}^{-}\right\rangle . \label{eq:spin-photon-corr}
\end{align}
Here, we have replaced the sum over spins $\sum_k$ with the factor $N$, and the sum of the spins coupled to any partiular spin $\sum_{k'\neq k}$ with the factor $N-1$.  The population difference $\left\langle \hat{\sigma}_{k}^{z}\right\rangle $ and the spin-spin correlation $\left\langle \hat{\sigma}_{k}^{\dagger}\hat{\sigma}_{k'}^{-}\right\rangle $ still follow  Eq. \eqref{eq:inversion} and Eq. \eqref{eq:spinspincorrelation},
respectively. 

The photon-photon correlation $\left\langle \hat{a}^{\dagger}\hat{b}\right\rangle $
follows the equation 
\begin{align}
 & \frac{\partial}{\partial t}\left\langle \hat{a}^{\dagger}\hat{b}\right\rangle =i\left(\tilde{\omega}_{c}^{*}-\tilde{\omega}_{f}\right)\left\langle \hat{a}^{\dagger}\hat{b}\right\rangle   +iNg_{k}\left\langle \hat{\sigma}_{k}^{\dagger}\hat{b}\right\rangle \nonumber \\
 & +iG\left(\left\langle \hat{b}^{\dagger}\hat{b}\right\rangle -\left\langle \hat{a}^{\dagger}\hat{a}\right\rangle \right). \label{eq:apb-new}
\end{align}
The mean photon number $\left\langle \hat{b}^{\dagger}\hat{b}\right\rangle $ in the
filter cavity and the spin-photon correlation $\left\langle \hat{\sigma}_{k}^{\dagger}\hat{b}\right\rangle $
still obey Eq.(4) in the main text and Eq. \eqref{eq:spinfilterphoton},
respectively. 

\begin{figure}[htbp]
\begin{centering}
\includegraphics[scale=0.45]{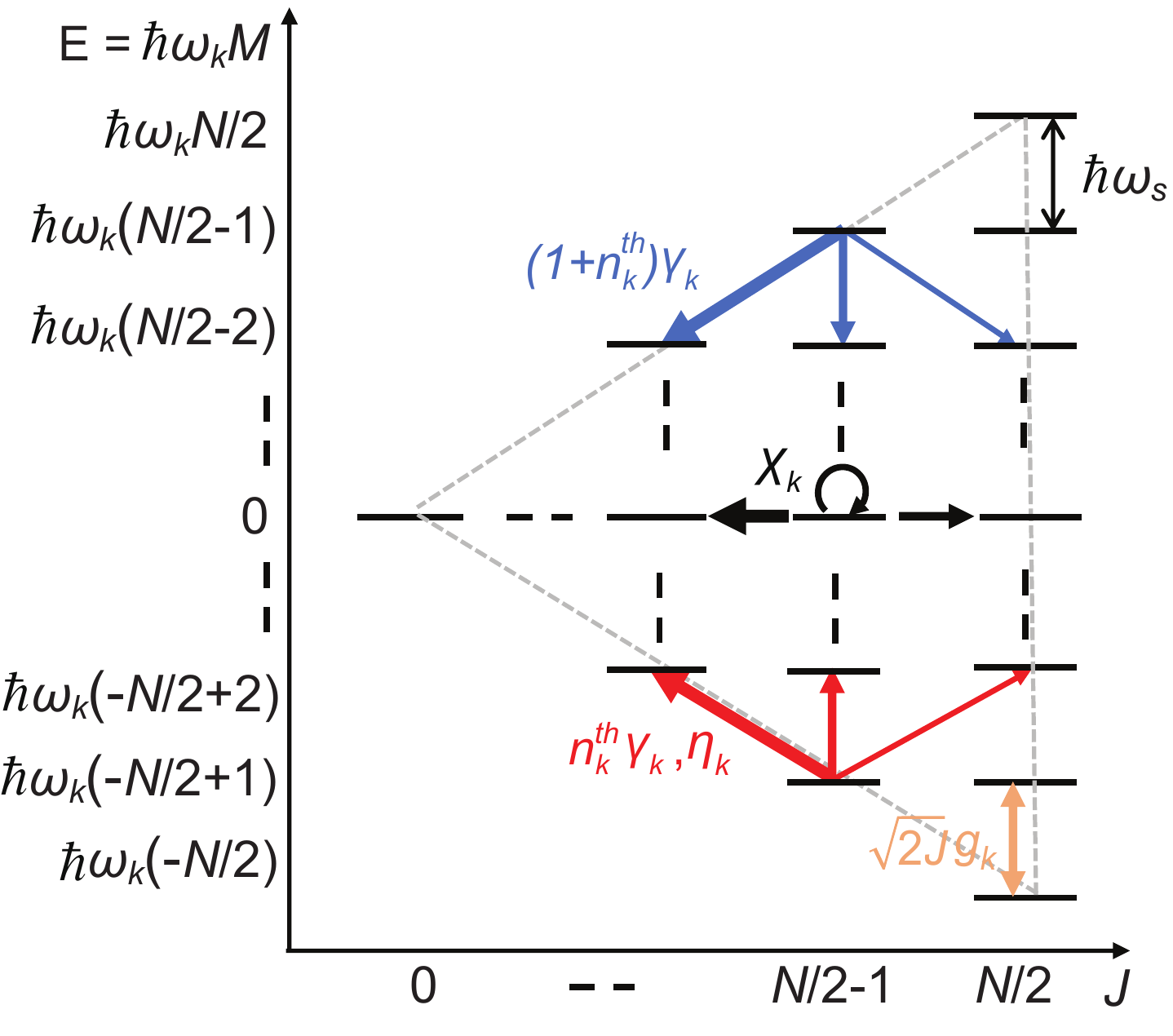}
\par\end{centering}
\caption{\label{fig:QumJumps} Relative probability (thickness of arrows) of the quantum jumps caused by the spin relaxation with rate $(1+n_k^{th})\gamma_k$ (blue arrows), and the spin thermal or external pumping with rate $n_k^{th}\gamma_k, \eta_k$ (red arrows), as well as the spin dephasing with rate $\chi_k$ (black arrows), among the Dicke states (black horizontal lines, gray dotted lines for the boundary). The coherent and collective coupling with the resonator mode is shown by the orange arrow, and the coupling strength depends on $J$ and $M$. Note that incoherent quantum jumps occur between two adjacent Dicke states with same or different $J$, while the coherent coupling occurs only  between states with the same $J$.  Here, the energy levels are for a spin-ensemble with an even number of spins. Energy levels for an odd number of spins are similar except that there are two levels for $M=\pm1/2$.}
\end{figure}

\section{Quantum Jumps and Coherent Coupling among Dicke States }{\label{sec:QumJumps}}

To understand the system dynamics in the  Dicke state representation, $\left| J,M \right \rangle$, we separately address the incoherent quantum jumps \citep{YZhang18,NShammah} and the coherent coupling. The Dicke states are characterized by two integer or half-integer numbers $J=N/2,...,0$ and $M=-J,...J$, and their energy levels are normally arranged in the way as shown in Fig. \ref{fig:QumJumps}. Note that the levels for given $J$ form a ladder with the spacing $\hbar\omega_k$, starting from $-J\hbar\omega_k$ and ending at $J\hbar\omega_k$. For convenience, the ladders for different $J$ are shifted horizontally in the figure to form a triangular pattern (grey dashed lines). 

For the spin relaxation (blue arrows) with the rate $(1+n_k^{th})\gamma_k$, the jumps occur towards the states with reduced $M$, and are dominated by the one 
towards the states with a smaller value of $J$ (with large probability). Note that for the Dicke states with $M=-J$ (lowest rung of the Dicke ladders) jumps are only possible to the states with larger $J$ due to the absence of states with $M<-J$. The upward quantum jumps for the spin pumping (red arrows) with rate $n_s^{th}\gamma_k, \eta_k$ (thermal pumping or external pumping) are simply the vertical mirror of those for the spin relaxation process, while those for dephasing with rate $\chi_k$ occur among states with the same value of $M$. Coherent coupling with the resonator mode introduces reversible vertical transitions between adjacent Dicke states of same $J$ (orange double-head arrow), with coupling strengths that increases with increasing $J$ and decreasing $|M|$. 

If only the spin relaxation occurs, the dynamics for the initially fully excited spin-ensemble  $\left| J=N/2,M=J \right \rangle$ follows first the upmost rung of Dicke ladders $\left| J,M\approx J \right \rangle$ and then the lowest rung $\left| J,M\approx -J \right \rangle$, and ends up at the ground state $\left| J=N/2,M=-J \right \rangle$. If the spin-ensemble is initially in the ground state and subject only to incoherent excitation, this dynamics is merely reversed. If subject to only dephasing, the spin-ensemble in an initial Dicke state $\left| J,M \right \rangle$ follows the horizontal arrows towards the leftmost rung of Dicke ladders ($\left| J=|M|,M\right \rangle$) with the same $M$. If subject only to the coherent coupling, the spin-ensemble initially in the ground state $\left| J=N/2,M=-J \right \rangle$, oscillates  up and down along the ladder of states with $J=N/2$. 

In the presence of all the processes, the dynamics becomes more complex and needs to be analyzed case by case. The simultaneous thermal-induced decay and excitation tend to cancel vertical motion, and their combined effect is similar to the quantum jumps of the spin dephasing, ending up, however, at a non-inverted equilibrium excitation at the lower rung of the Dicke ladders with $J<N/2$ (see Fig. 4 in the main text). When the external pumping $\eta_k$ is included, the quantum jumps due to the spin-pumping are enhanced and can overcome those due to the thermal decay, and the spin-ensemble evolves towards the  upper rung of the Dicke ladders with $J=M>0$ (see Fig. 4 in the main text). The tendency to occupy the upmost and lowest rung of the Dicke ladders is associated with the higher combinatorial degeneracy of these states and is further enforced by the dephasing process. 

When adding the coherent coupling with the resonator mode, the system explores also the states in the middle of the Dicke ladders. For weak external pumping, the spin-ensemble is not population-inverted, and there are only few photons inside the resonator. As a result, the coherent coupling is relatively weak, and the dynamics is not affected so much. In contrast, for strong external pumping, the spin-ensemble becomes population-inverted, and the photon number inside the resonator increases dramatically due to the stimulated emission. The increase of the photon number increases also the stimulated absorption, which  tends to balance the stimulated emission, leading to significant population of the middle of Dicke ladders. However, in the presence of large dephasing, the system finally occupies the upper rung of Dicke ladders with $J/N\approx M/N\approx 0$, see Fig. 4 in the main text. Thus, except situations extremely near to the phase transition, the condition $|M|\approx J$ is satisfied in most cases, and the Holstein-Primakoff approximation discussed in Sec. \ref{sec:HolPriApp} can be applied.  

\section{Derivation of Transition Boundaries}{\label{sec:traBou}}

In the main text, we have observed the transitions between superradiance, superradiant maser and thermal regime. To understand the conditions leading to these transitions, in this section, we derive analytical expressions for these conditions. To this end, we focus on the stimulated processes and approximate Eq. \eqref{eq:spin-photon-corr}  as 
\begin{align}
 & \frac{\partial}{\partial t}\left\langle \hat{\sigma}_{k}^{\dagger}\hat{a}\right\rangle  \approx i\left(\tilde{\omega}_{k}^{*}-\tilde{\omega}_{c}\right)\left\langle \hat{\sigma}_{k}^{\dagger}\hat{a}\right\rangle -ig_{k}\left\langle \hat{a}^{\dagger}\hat{a}\right\rangle \left\langle \hat{\sigma}_{k}^{z}\right\rangle. \label{eq:spin-photon}
\end{align}
Next, we consider the steady-state solution of the above equation
$ \left\langle \hat{\sigma}_{k}^{\dagger}\hat{a}\right\rangle\approx  \frac{g_{k}}{ \tilde{\omega}_{k}^{*}-\tilde{\omega}_{c} }\left\langle \hat{a}^{\dagger}\hat{a}\right\rangle \left\langle \hat{\sigma}_{k}^{z}\right\rangle$, and obtain  
\begin{equation}
\left\langle \hat{\sigma}_{k}^{\dagger}\hat{a}\right\rangle - \left\langle a^{\dagger} \hat{\sigma}_{k}^{-}\right\rangle  \approx -i  \frac{k_{EET}}{g_{k}}   \left\langle \hat{a}^{\dagger}\hat{a}\right\rangle \left\langle \hat{\sigma}_{k}^{z}\right\rangle , \label{eq:comb}
\end{equation}
where we have introduced the energy transfer rate $ k_{EET} = \frac{  2 g_k^2 (\lambda_k^s + \kappa_c/2)}{ ( \omega_{k} - \omega_{c})^2 + (\lambda_k^s + \kappa_c/2)^2 } $.

Inserting Eq. \eqref{eq:comb} into the steady-state version of Eq. \eqref{eq:inversion}, we obtain 
\begin{equation}
 \left\langle \hat{\sigma}_{k}^{z}\right\rangle  \approx    \frac{\eta_{k}-\gamma_{k}}{2 k_{EET}   \left\langle \hat{a}^{\dagger}\hat{a}\right\rangle   +  \gamma_{k} \left(2n_{k}^{th}+1\right)+\eta_{k} }.\label{eq:inv_st}
\end{equation}
Inserting Eq. \eqref{eq:comb} in the steady-state version of Eq. \eqref{eq:inv_st}, we obtain 
\begin{equation}
 0 =-\kappa_c \left\langle \hat{a}^{\dagger}\hat{a}\right\rangle +\kappa_c n_{c}^{th} 
+N k_{EET}   \left\langle \hat{a}^{\dagger}\hat{a}\right\rangle  \left\langle \hat{\sigma}_{k}^{z}\right\rangle.
 \end{equation}
Using Eq. \eqref{eq:inv_st}, we can rewrite the above equation as $A \left\langle \hat{a}^{\dagger}\hat{a}\right\rangle^2 + B \left\langle \hat{a}^{\dagger}\hat{a}\right\rangle + C =0 $ with $A=2k_{EET}$, $B= \gamma_{k} \left(2n_{k}^{th}+1\right)+\eta_{k}  - [2 n_c^{th} + N(\eta_k-\gamma_k)/\kappa_c]k_{EET}$, $C= - n_c^{th} [(\gamma_{k} \left(2n_{k}^{th}+1\right)+\eta_{k} ]$. For our systems, $B^2 \gg AC$, we obtain the solution $ \left\langle \hat{a}^{\dagger}\hat{a}\right\rangle = - B/A$. Assuming that the stimulated processes dominate over the thermal process $\left\langle \hat{a}^{\dagger}\hat{a}\right\rangle \geq n_c^{th}$, we obtain the condition for the incoherent pumping
 \begin{equation}
 \frac{\eta_k}{\gamma_k} \geq \frac{2n_{k}^{th}+1+N\mathcal{C}}{N\mathcal{C}-1}, \end{equation}  where we have introduced the single-particle cooperativity $\mathcal{C} = k_{EET}/\kappa_c$.   
 
 \section{Semi-analytical Expression for Spectral Peak Positions and Linewidths }{\label{sec:spelw}}
 
 In this appendix, we derive the semi-analytical expressions for the spectral peak positions and linewidths. To this end, we consider the steady-state solution of
Eq. \eqref{eq:spinfilterphoton} :
\begin{equation}
\left\langle \sigma_{k}^{\dagger}\hat{b}\right\rangle =  \tilde{\omega}_{kf}^{-1}\left(g_{k}\left\langle \hat{a}^{\dagger}\hat{b}\right\rangle \left\langle \hat{\sigma}_{k}^{z}\right\rangle +G\left\langle \hat{\sigma}_{k}^{\dagger}\hat{a}\right\rangle \right), 
\end{equation} with the complex frequency $\tilde{\omega}_{kf}=\omega_{k}-\omega_{f}+i\left(\lambda_{k}^s+\kappa_{f}/2\right)$. 
Inserting the above expression into Eq. \eqref{eq:apb-new}, we get the steady-state photon-photon correlation
\begin{equation}
\left\langle \hat{a}^{\dagger}\hat{b}\right\rangle =G\frac{\left\langle \hat{a}^{\dagger}\hat{a}\right\rangle -\left\langle \hat{b}^{\dagger}\hat{b}\right\rangle - \tilde{\omega}_{ kf}^{-1}N g_{k}\left\langle \sigma_{ k}^{\dagger}\hat{a}\right\rangle }{\tilde{\omega}_{cf}+ \tilde{\omega}_{kf}^{-1}N g_{k}^{2}\left\langle \hat{\sigma}_{k}^{z}\right\rangle }, 
\end{equation}
with the complex frequency $\tilde{\omega}_{cf}=\omega_{c}-\omega_{f}+i\frac{1}{2}\left(\kappa_{c}+\kappa_{f}\right)$.  Inserting the above results into Eq.(4) of the main text, we get the steady-state mean photon number in the filter cavity 
\begin{equation}
\left\langle \hat{b}^{\dagger}\hat{b}\right\rangle =\frac{G^{2}2\mathrm{Im}\left[\frac{\tilde{\omega}_{ kf} \left\langle \hat{a}^{\dagger}\hat{a}\right\rangle -  N g_{k}\left\langle \hat{\sigma}_{ k}^{\dagger}\hat{a}\right\rangle }{\tilde{\omega}_{ kf}\tilde{\omega}_{cf}+  N g_{k}^{2}\left\langle \hat{\sigma}_{k}^{z}\right\rangle }\right]}{G^{2}2\mathrm{Im}\left[\tilde{\omega}_{cf}+ \omega_{kf}^{-1}N g_{k}^{2}\left\langle \hat{\sigma}_{k}^{z}\right\rangle \right]^{-1}-\kappa_{f}}.\label{eq:bpb-sem-ana}
\end{equation}
To resolve the spectrum with the filter resonator approach, we require  $\kappa_f$ to be smaller than the spectral feature to be resolved, and  $G$ to be small enough to reduce the backaction on the main resonator. By analyzing the numerator and denominator of Eq. \eqref{eq:bpb-sem-ana},  we find that the spectral peak positions and linewidth are mainly determined by the denominator. Assuming that the denominator can be written as the product  $\tilde{\omega}_{kf}\tilde{\omega}_{cf} + N g_{k}^{2}\left\langle \hat{\sigma}_{k}^{z}\right\rangle \approx (\omega_f -\tilde{\omega}_+ )(\omega_f - \tilde{\omega}_- ) $, we obtain two complex frequencies $\tilde{\omega}_\pm = [\tilde{\omega}_k^* + \tilde{\omega}_c^* \pm \sqrt{(\tilde{\omega}_k^* -\tilde{\omega}_c^*)^2 - 4Ng_k^2  \left\langle \hat{\sigma}_{k}^{z}\right\rangle } ]/2$.

For the weak pumping,  we have $\left\langle \hat{\sigma}_{k}^{z}\right\rangle \approx 2M/N \approx -2J/N<0$ and thus the solutions become  $\tilde{\omega}_\pm = [\tilde{\omega}_k^* + \tilde{\omega}_c^* \pm \sqrt{(\tilde{\omega}_k-\tilde{\omega}_c)^2 + 8 g_k^2 J } ]/2$. These solutions result in two peaks in the spectra with frequencies and linewidths given by  their real and imaginary parts. For strong pumping, we have the population inversion $\left\langle \hat{\sigma}_{k}^{z}\right\rangle>0$, and in this case, one of the solutions has a smaller linewidth, and is responsible for masing. 

In particular, for the resonant condition $\omega_c =\omega_k$, we obtain $\tilde{\omega}_\pm = \omega_c + i (\lambda_k^s + \kappa_c/2 \pm \sqrt{R} )/2$ with the abbreviation  $R = (\lambda_k^s -\kappa_c/2)^2 + 4Ng_k^2\left\langle \hat{\sigma}_{k}^{z}\right\rangle$. For $R <0 $ or $\left\langle \hat{\sigma}_{k}^{z}\right\rangle <- (\lambda_k^s -\kappa_c/2 )^2/(4Ng_k^2)$,  we obtain  $\tilde{\omega}_\pm = \omega_c\mp \sqrt{|R|}/2 +i (\lambda_k^s + \kappa_c/2)/2$.  As a result, the spectrum shows two peaks centered around $\omega_c\mp \sqrt{|R|}/2$ (separated by $\sqrt{|R|}$)  with the linewidth $ \lambda_k^s + \kappa_c/2$. When  $\left\langle \hat{\sigma}_{k}^{z}\right\rangle$ approaches zero for increasing pumping,  $\sqrt{|R|}$ decreases and thus the two peaks approach each other and finally merge into one. When $R >0 $, $\tilde{\omega}_\pm$ have the same real part $\omega_c $, but different imaginary part $(\lambda_k^s + \kappa_c/2 \pm \sqrt{R} )/2$. As a result, the two peaks center around  $\omega_c $, and one peak has larger linewidth  $(\lambda_k^s + \kappa_c/2 +\sqrt{R} )/2$ and the other one has smaller linewidth $(\lambda_k^s + \kappa_c/2 - \sqrt{R} )/2$ (corresponds to the masing).

\begin{figure}[htbp]
\begin{centering}
\includegraphics[scale=0.5]{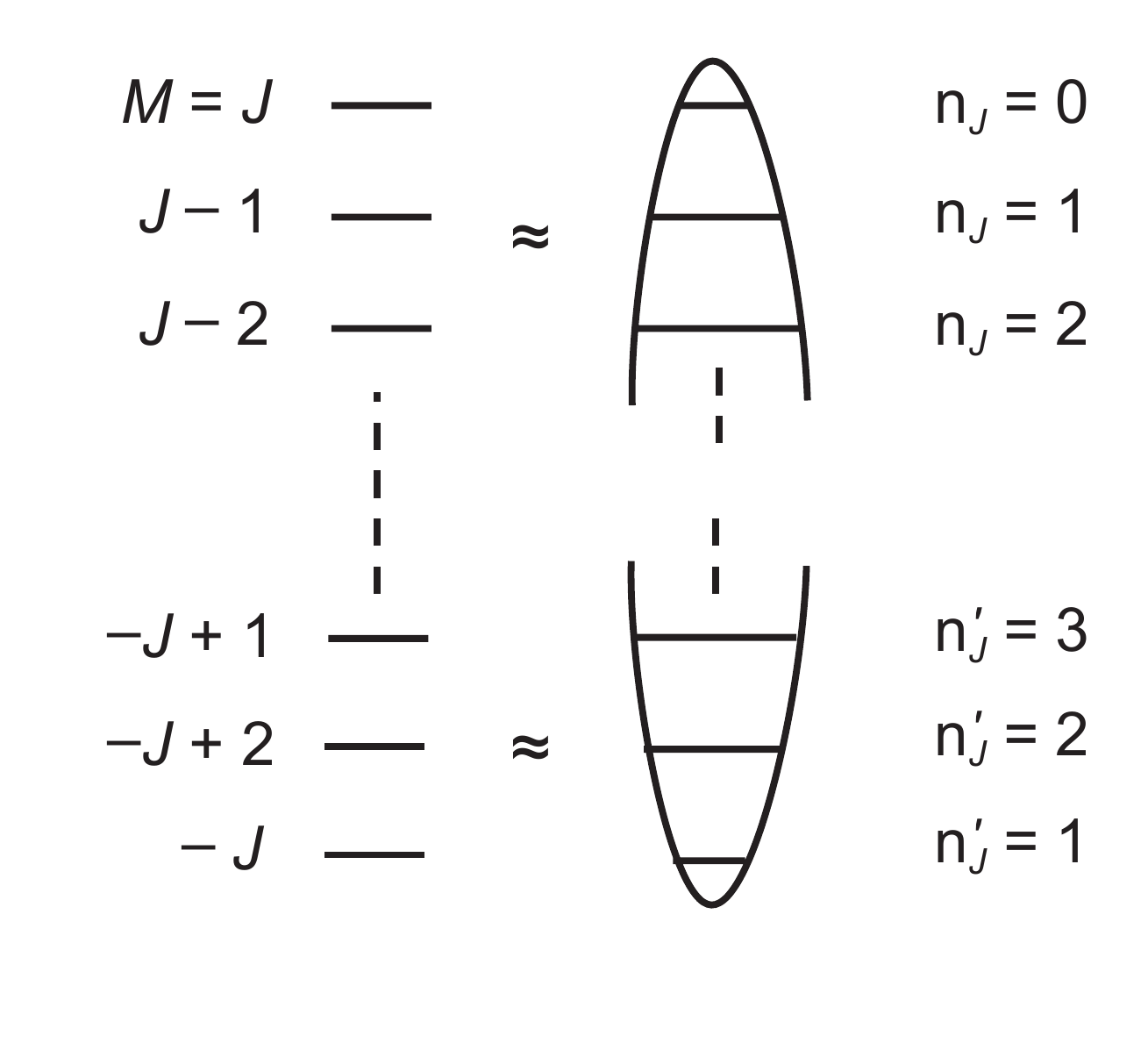}
\par\end{centering}
\caption{\label{fig:HolPri}  Approximation of Dicke states near the top and bottom of a Dicke ladder for given $J$ (left) with the occupation number states of the upside-down and normal quantized harmonic oscillator (right). }
\end{figure}

\section{ Holstein-Primakoff Approximation}{\label{sec:HolPriApp}}

In the main text, we have observed that the spin-ensemble  occupies the upper and lower rung of Dicke ladders for strong and weak pumping, which is caused by the interplay of quantum jumps of the spin relaxation, spin pumping, spin dephasing and the coherent coupling with the resonator, as discussed in App. \ref{sec:QumJumps}. Here we show that these situations allow us to derive approximate Hamiltonians for the spin ensemble-resonator coupling to understand the physics leading to the masing and the double peak spectrum, respectively. To this end, we consider the spin Hamiltonian $\hat{H}_s = \left(\hbar\omega_{s}/2\right)  \sum_{k=1}^{N}\hat{\sigma}_{k}^{z} $  and the spin-resonator interaction $\hat{H}_{s-c} = \hbar g_s\left( \sum_{k} \hat{\sigma}_{k}^{\dagger}\hat{a}+a^{\dagger} \sum_{k} \hat{\sigma}_{k}^{-}\right)$ for the spins with identical frequency $\omega_{s}$ and identical coupling with the resonator $g_s$. 

\subsection{Parametric Coupling \label{subsec:ParCoup}}

To proceed, we apply the Holstein-Primakoff (HP) transformation to represent the Dicke states $\left| J,M \right\rangle$ to the occupation number states $\left| n_J \right\rangle$ of a quantized harmonic oscillator characterized by the creation $\hat{b}^\dagger_J$ and annihilation operator $\hat{b}_J$. The normal transformation assumes  the relation $M=J-n_J$, which leads to the mapping $\left|J,M\pm1\right\rangle \to \left|J,n_J \mp1\right\rangle$ \citep{JAGyamfi,THolstein} and thus to the relationships $\sum_k \sigma^z_k = 2\sum_J (J-\hat{b}_J^\dagger\hat{b}_J)$, $\sum_k \sigma^\dagger_k = \sum_J \sqrt{2J-\hat{b}^\dagger_J\hat{b}_J} \hat{b}_J $, $\sum_k \sigma^-_k = \sum_J \hat{b}^\dagger_J \sqrt{2J-\hat{b}^\dagger_J \hat{b}_J}$. In this case, the vacuum state $\left| n_J=0 \right\rangle$ is associated with the upper Dicke state $\left| J,J \right\rangle$ for given $J$. For the Dicke states with $n_J \ll 2J$, we have $\sqrt{2J-\hat{b}^\dagger_J \hat{b}_J}\left| n_J \right\rangle =\sqrt{2J-n_J}\left| n_J \right\rangle \approx \sqrt{2J}\left| n_J \right\rangle$ and thus  $\sum_k \sigma^\dagger_k \approx \sum_J \sqrt{2J} \hat{b}_J $, $\sum_k \sigma^-_k \approx \sum_J \hat{b}^\dagger_J \sqrt{2J}$, which approximate effectively these Dicke states as the occupation number states of an inverted quantized harmonic oscillator (upper part of Fig. \ref{fig:HolPri}). Using these two relations, we can rewrite the spin-resonator interaction as the parametric coupling 
\begin{equation}
\hat{H}_{s-c} \approx \hbar \sum_J\sqrt{2J}g_s\left( \hat{b}_J \hat{a}+\hat{a}^{\dagger}\hat{b}_J^\dagger\right). \label{eq:ParCou} 
\end{equation}
In our system, the spin-ensemble occupies the upper rung of Dicke ladders for strong pumping, and thus the masing for strong pumping is well described by the parametric coupling of Eq. \eqref{eq:ParCou}. The parametric coupling leads to the gain and to spin-photon entanglement (two-mode squeezing) \citep{KDebnath}.

\begin{figure}[htbp]
\begin{centering}
\includegraphics[scale=0.5]{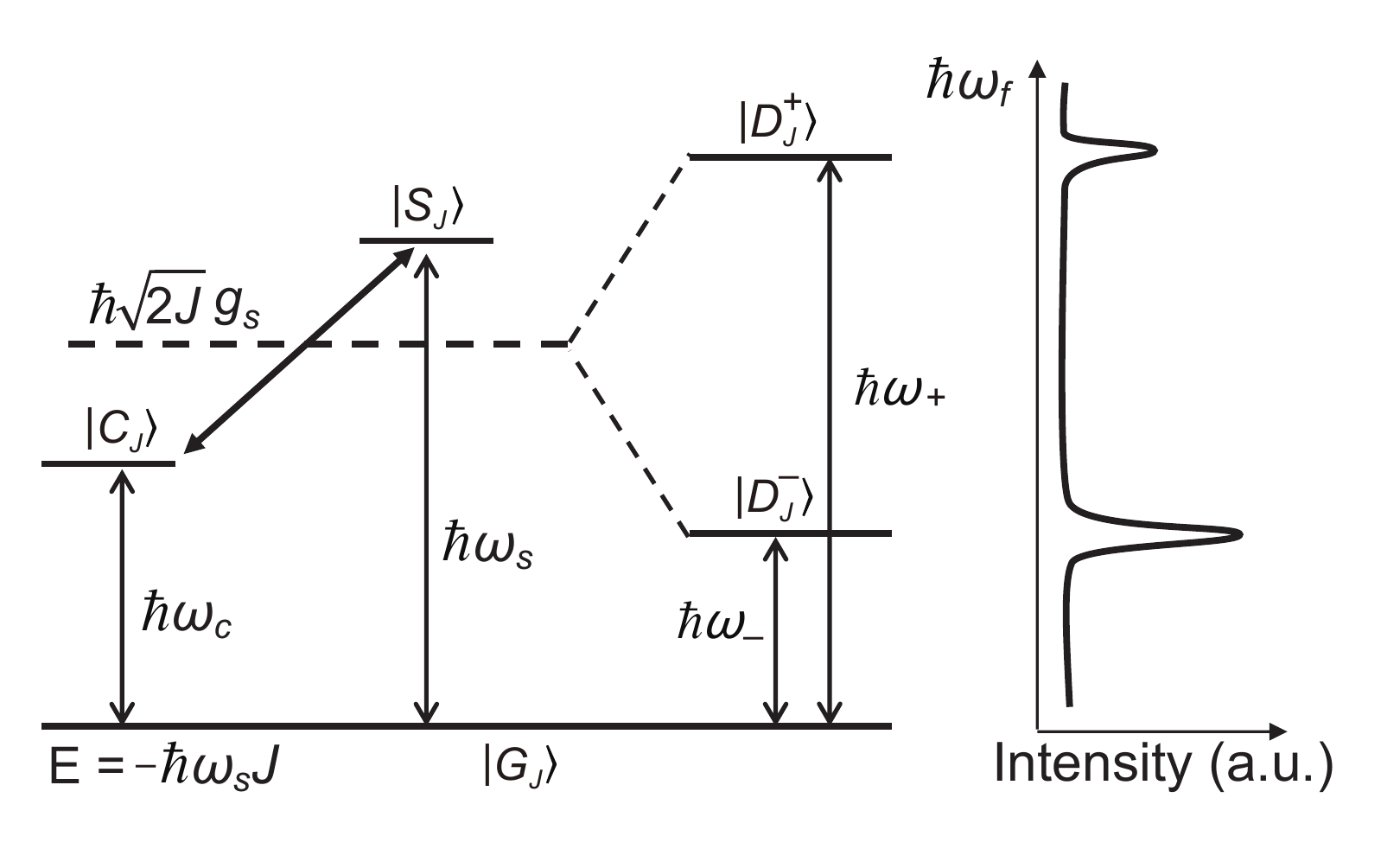}
\par\end{centering}
\caption{\label{fig:SpiPhoDreStates} Formation of spin-photon dressed states. Left part shows the singly excited  states  $\left | C_J \right \rangle$ (with a single photon) and $\left | S_J \right \rangle $ (with a single spin excitation in the ensemble) with the mutual coupling $\hbar \sqrt{2J}g_s$ for given $J$. The  middle part of the figure shows the dressed eigenstates states  $\left | D_J^{+} \right \rangle$, $\left | D_J^{-} \right \rangle$, leading to the double peaks in the emission spectrum.  }
\end{figure}

\subsection{Spin-Photon Dressed States \label{subsec:DreStates}}
We can also apply the HP transformation $M=n'_J-J$ and the mapping $\left|J,M\pm 1\right\rangle \to \left|J,n'_J\pm1\right\rangle$, which lead to the relations $\sum_k \hat{\sigma}^z_k =  2\sum_J (\hat{b}_J^\dagger\hat{b}_J-J)$, $\sum_k \hat{\sigma}^\dagger_k 	= \hat{b}_{J}^{\dagger}\sqrt{2J-\hat{b}_{J}^{\dagger} \hat{b}_{J}}$,$\sum_k \hat{\sigma}^-_k =\sqrt{2J-\hat{b}_{J}^{\dagger}\hat{b}_{J}}\hat{b}_{J}$. In this case, the vacuum state $\left| n'_J=0 \right\rangle$ is associated with the lowest Dicke state  $\left| J,-J \right\rangle$ for any given $J$. For the Dicke states with $n'_J \ll 2J$, we have $\sum_k \hat{\sigma}^\dagger_k \approx \hat{b}_{J}^{\dagger}\sqrt{2J}$,$\sum_k \hat{\sigma}^-_k \approx \sqrt{2J} \hat{b}_{J}$. Using these two relations, we can rewrite the spin-resonator interaction as the coupling of two harmonic oscillators 
\begin{equation}
\hat{H}_{s-c} \approx \hbar \sum_J\sqrt{2J}g_s\left( \hat{b}_J^\dagger \hat{a}+\hat{a}^{\dagger}\hat{b}_J\right). \label{eq:HscApp}
\end{equation}

In the low excitation limit, we can introduce the ground product state $\left|G_J\right\rangle = \left|n=0\right\rangle \left|n'_J=0\right\rangle$, and the singly excited states $\left|C_J\right\rangle = \left|n=1\right\rangle \left|n'_J=0\right\rangle$, $\left|S_J\right\rangle= \left|n=0\right\rangle \left|n'_J=1\right\rangle$, where $\left|n=0\right\rangle,\left|n=1\right\rangle$
are the vacuum state and single-photon state, and the labeling $S,C$ distinguish the singly excited states with the excited spin-ensemble or the excited photon. Using these states, we can approximate $\hat{b}_J^\dagger \hat{a}\approx \left|S_J\right\rangle \left\langle C_J\right|$ and $\hat{a}^\dagger\hat{b}_J \approx \left|C_J\right\rangle \left\langle S_J\right|$, and thus obtain $\hat{H}_{s-c} \approx \hbar \sum_J\sqrt{2J}g_s\left( \left|S_J\right\rangle \left\langle C_J\right| + \left|C_J\right\rangle \left\langle S_J\right|\right)$. Furthermore, we can approximate the sum of the resonator Hamiltonian and the spin-ensemble Hamiltonian as $\hat{H}_c + \hat{H}_s \approx \hbar \sum_J [(\omega_c -\omega_sJ) \left|C_J\right\rangle\left\langle C_J\right| - \omega_s (J-1) \left|S_J\right\rangle\left\langle S_J\right| - \omega_s J \left|G_J\right\rangle\left\langle G_J\right|]$. From these Hamiltonians we can obtain the energy levels and the coupling (for given $J$) as shown on the left side of Fig. \ref{fig:SpiPhoDreStates}.

After diagonalizing the approximate Hamiltonians for given $J$, we obtain the eigen-frequencies $d_\pm = \frac{1}{2}[\omega_c+\omega_s - 2J\omega_s \pm \sqrt{8Jg_s^2 + (\omega_c-\omega_s)^2}]$ for the dressed spin-photon states $\left|D^\pm_J\right\rangle =\frac{1}{N_{\pm}}[\omega_c-\omega_s \pm \sqrt{8Jg_s^2 + (\omega_c-\omega_s)^2}]\left|C_J\right\rangle + (2\sqrt{2J}g_s)\left|S_J\right\rangle$ with the normalization factors $N_{\pm}=\sqrt{16Jg_s^2 \pm (\omega_c-\omega_s)\sqrt{8Jg_s^2+(\omega_c-\omega_s)^2}}$. The transition frequency between the dressed states and the ground state for given $J$ can be readily calculated as $\omega_\pm = d_\pm - (-\omega_sJ)=\frac{1}{2}[\omega_c+\omega_s \pm \sqrt{8Jg_s^2 + (\omega_c-\omega_s)^2}]$. The energy levels of the dressed states are shown in the middle of Fig. \ref{fig:SpiPhoDreStates}, and the transitions between them and the ground product state are responsible for the double-peak spectrum in the system with weak pumping, see the right side of Fig. \ref{fig:SpiPhoDreStates}. Furthermore, we note that $\omega_\pm$ are consistent with the real parts of $\tilde{\omega}_\pm$ as identified in Sec. \ref{sec:spelw}. Note the relationship $J \approx -M =-\frac{1}{2}N\left\langle \hat{\sigma}_k^z \right\rangle$ for the weak pumping in our system.

\end{document}